\documentclass[12pt]{article}

\usepackage{graphicx}
\usepackage{comment}
\usepackage{amsmath}
\usepackage{float}
\usepackage{longtable}
\usepackage{caption}
\usepackage{hyperref}
\usepackage{xcolor}

\input{epsf}

\advance \topmargin by -\headheight
\advance \topmargin by -\headsep

\evensidemargin \oddsidemargin
\begin{document}

\topmargin -.6in

\def\rh{{\hat \rho}}
\def\alie{{\hat{\cal G}}}
\newcommand{\sect}[1]{\setcounter{equation}{0}\section{#1}}
\renewcommand{\theequation}{\thesection.\arabic{equation}}

\def\rf#1{(\ref{eq:#1})}
\def\lab#1{\label{eq:#1}}
\def\nonu{\nonumber}
\def\br{\begin{eqnarray}}
\def\er{\end{eqnarray}}
\def\be{\begin{equation}}
\def\ee{\end{equation}}
\def\eq{\!\!\!\! &=& \!\!\!\! }
\def\foot#1{\footnotemark\footnotetext{#1}}
\def\lb{\lbrack}
\def\rb{\rbrack}
\def\llangle{\left\langle}
\def\rrangle{\right\rangle}
\def\blangle{\Bigl\langle}
\def\brangle{\Bigr\rangle}
\def\llbrack{\left\lbrack}
\def\rrbrack{\right\rbrack}
\def\lcurl{\left\{}
\def\rcurl{\right\}}
\def\({\left(}
\def\){\right)}
\newcommand{\nit}{\noindent}
\newcommand{\ct}[1]{\cite{#1}}
\newcommand{\bi}[1]{\bibitem{#1}}
\def\lskip{\vskip\baselineskip\vskip-\parskip\noindent}
\relax

\def\hpsi{\widehat{\psi}}

\def\tr{\mathop{\rm Tr}}
\def\Tr{\mathop{\rm Tr}}
\def\trace{\widehat{\rm Tr}}
\def\v{\vert}
\def\bv{\bigm\vert}
\def\Bgv{\;\Bigg\vert}
\def\bgv{\bigg\vert}
\newcommand\partder[2]{{{\partial {#1}}\over{\partial {#2}}}}
\newcommand\funcder[2]{{{\delta {#1}}\over{\delta {#2}}}}
\newcommand\Bil[2]{\Bigl\langle {#1} \Bigg\vert {#2} \Bigr\rangle}  
\newcommand\bil[2]{\left\langle {#1} \bigg\vert {#2} \right\rangle} 
\newcommand\me[2]{\left\langle {#1}\bv {#2} \right\rangle} 
\newcommand\sbr[2]{\left\lbrack\,{#1}\, ,\,{#2}\,\right\rbrack}
\newcommand\pbr[2]{\{\,{#1}\, ,\,{#2}\,\}}
\newcommand\pbbr[2]{\lcurl\,{#1}\, ,\,{#2}\,\rcurl}

\def\ket#1{\mid {#1} \rangle}
\def\bra#1{\langle {#1} \mid}
\newcommand{\braket}[2]{\langle {#1} \mid {#2}\rangle}
%
\def\a{\alpha}
\def\at{{\tilde A}^R}
\def\atc{{\tilde {\cal A}}^R}
\def\atcm#1{{\tilde {\cal A}}^{(R,#1)}}
\def\b{\beta}
\def\dc{{\cal D}}
\def\d{\delta}
\def\D{\Delta}
\def\eps{\epsilon}
\def\vareps{\varepsilon}
\def\g{\gamma}
\def\G{\Gamma}
\def\grad{\nabla}
\def\h{{1\over 2}}
\def\l{\lambda}
\def\L{\Lambda}
\def\m{\mu}
\def\n{\nu}
\def\o{\over}
\def\om{\omega}
\def\O{\Omega}
\def\p{\phi}
\def\P{\Phi}
\def\pa{\partial}
\def\pr{\prime}
\def\pt{{\tilde \Phi}}
\def\qs{Q_{\bf s}}
\def\ra{\rightarrow}
\def\s{\sigma}
\def\S{\Sigma}
\def\t{\tau}
\def\th{\theta}
\def\Th{\Theta}
\def\tpp{\Theta_{+}}
\def\tmm{\Theta_{-}}
\def\tpg{\Theta_{+}^{>}}
\def\tms{\Theta_{-}^{<}}
\def\tp0{\Theta_{+}^{(0)}}
\def\tm0{\Theta_{-}^{(0)}}
\def\ti{\tilde}
\def\wti{\widetilde}
\def\jc{J^C}
\def\bj{{\bar J}}
\def\sj{{\jmath}}
\def\bsj{{\bar \jmath}}
\def\bp{{\bar \p}}
\def\vp{\varphi}
\def\ve{\varepsilon}
\def\vt{{\tilde \varphi}}
\def\faa{Fa\'a di Bruno~}
\def\ca{{\cal A}}
\def\cb{{\cal B}}
\def\ce{{\cal E}}
\def\cg{{\cal G}}
\def\cgh{{\hat {\cal G}}}
\def\ch{{\cal H}}
\def\chh{{\hat {\cal H}}}
\def\cl{{\cal L}}
\def\cm{{\cal M}}
\def\cn{{\cal N}}
\def\u2{\mid u\mid^2}
\newcommand\sumi[1]{\sum_{#1}^{\infty}}   
\newcommand\fourmat[4]{\left(\begin{array}{cc}  
{#1} & {#2} \\ {#3} & {#4} \end{array} \right)}

%
\def\lie{{\cal G}}
\def\kmlie{{\hat{\cal G}}}
\def\dlie{{\cal G}^{\ast}}
\def\elie{{\widetilde \lie}}
\def\edlie{{\elie}^{\ast}}
\def\hlie{{\cal H}}
\def\flie{{\cal F}}
\def\wlie{{\widetilde \lie}}
\def\f#1#2#3 {f^{#1#2}_{#3}}
\def\winf{{\sf w_\infty}}
\def\win1{{\sf w_{1+\infty}}}
\def\hwinf{{\sf {\hat w}_{\infty}}}
\def\Winf{{\sf W_\infty}}
\def\Win1{{\sf W_{1+\infty}}}
\def\hWinf{{\sf {\hat W}_{\infty}}}
\def\Rm#1#2{r(\vec{#1},\vec{#2})}          
\def\OR#1{{\cal O}(R_{#1})}           
\def\ORti{{\cal O}({\widetilde R})}           
\def\AdR#1{Ad_{R_{#1}}}              
\def\dAdR#1{Ad_{R_{#1}^{\ast}}}      
\def\adR#1{ad_{R_{#1}^{\ast}}}       
\def\KP{${\rm \, KP\,}$}                 
\def\KPl{${\rm \,KP}_{\ell}\,$}         
\def\KPo{${\rm \,KP}_{\ell = 0}\,$}         
\def\mKPa{${\rm \,KP}_{\ell = 1}\,$}    
\def\mKPb{${\rm \,KP}_{\ell = 2}\,$}    
%
\def\rlx{\relax\leavevmode}
\def\inbar{\vrule height1.5ex width.4pt depth0pt}
\def\IZ{\rlx\hbox{\sf Z\kern-.4em Z}}
\def\IR{\rlx\hbox{\rm I\kern-.18em R}}
\def\IC{\rlx\hbox{\,$\inbar\kern-.3em{\rm C}$}}
\def\IN{\rlx\hbox{\rm I\kern-.18em N}}
\def\IO{\rlx\hbox{\,$\inbar\kern-.3em{\rm O}$}}
\def\IP{\rlx\hbox{\rm I\kern-.18em P}}
\def\IQ{\rlx\hbox{\,$\inbar\kern-.3em{\rm Q}$}}
\def\IF{\rlx\hbox{\rm I\kern-.18em F}}
\def\IG{\rlx\hbox{\,$\inbar\kern-.3em{\rm G}$}}
\def\IH{\rlx\hbox{\rm I\kern-.18em H}}
\def\II{\rlx\hbox{\rm I\kern-.18em I}}
\def\IK{\rlx\hbox{\rm I\kern-.18em K}}
\def\IL{\rlx\hbox{\rm I\kern-.18em L}}
\def\one{\hbox{{1}\kern-.25em\hbox{l}}}
\def\0#1{\relax\ifmmode\mathaccent"7017{#1}%
B        \else\accent23#1\relax\fi}
\def\omz{\0 \omega}
%
\def\ltimes{\mathrel{\vrule height1ex}\joinrel\mathrel\times}
\def\rtimes{\mathrel\times\joinrel\mathrel{\vrule height1ex}}
%
\def\mark{\noindent{\bf Remark.}\quad}
\def\prop{\noindent{\bf Proposition.}\quad}
\def\theor{\noindent{\bf Theorem.}\quad}
\def\name{\noindent{\bf Definition.}\quad}
\def\exam{\noindent{\bf Example.}\quad}
\def\proof{\noindent{\bf Proof.}\quad}

%
%
\def\PRL#1#2#3{{\sl Phys. Rev. Lett.} {\bf#1} (#2) #3}
\def\NPB#1#2#3{{\sl Nucl. Phys.} {\bf B#1} (#2) #3}
\def\NPBFS#1#2#3#4{{\sl Nucl. Phys.} {\bf B#2} [FS#1] (#3) #4}
\def\CMP#1#2#3{{\sl Commun. Math. Phys.} {\bf #1} (#2) #3}
\def\PRD#1#2#3{{\sl Phys. Rev.} {\bf D#1} (#2) #3}
\def\PLA#1#2#3{{\sl Phys. Lett.} {\bf #1A} (#2) #3}
\def\PLB#1#2#3{{\sl Phys. Lett.} {\bf #1B} (#2) #3}
\def\JMP#1#2#3{{\sl J. Math. Phys.} {\bf #1} (#2) #3}
\def\PTP#1#2#3{{\sl Prog. Theor. Phys.} {\bf #1} (#2) #3}
\def\SPTP#1#2#3{{\sl Suppl. Prog. Theor. Phys.} {\bf #1} (#2) #3}
\def\AoP#1#2#3{{\sl Ann. of Phys.} {\bf #1} (#2) #3}
\def\PNAS#1#2#3{{\sl Proc. Natl. Acad. Sci. USA} {\bf #1} (#2) #3}
\def\RMP#1#2#3{{\sl Rev. Mod. Phys.} {\bf #1} (#2) #3}
\def\PR#1#2#3{{\sl Phys. Reports} {\bf #1} (#2) #3}
\def\AoM#1#2#3{{\sl Ann. of Math.} {\bf #1} (#2) #3}
\def\UMN#1#2#3{{\sl Usp. Mat. Nauk} {\bf #1} (#2) #3}
\def\FAP#1#2#3{{\sl Funkt. Anal. Prilozheniya} {\bf #1} (#2) #3}
\def\FAaIA#1#2#3{{\sl Functional Analysis and Its Application} {\bf #1} (#2)
#3}
\def\BAMS#1#2#3{{\sl Bull. Am. Math. Soc.} {\bf #1} (#2) #3}
\def\TAMS#1#2#3{{\sl Trans. Am. Math. Soc.} {\bf #1} (#2) #3}
\def\InvM#1#2#3{{\sl Invent. Math.} {\bf #1} (#2) #3}
\def\LMP#1#2#3{{\sl Letters in Math. Phys.} {\bf #1} (#2) #3}
\def\IJMPA#1#2#3{{\sl Int. J. Mod. Phys.} {\bf A#1} (#2) #3}
\def\AdM#1#2#3{{\sl Advances in Math.} {\bf #1} (#2) #3}
\def\RMaP#1#2#3{{\sl Reports on Math. Phys.} {\bf #1} (#2) #3}
\def\IJM#1#2#3{{\sl Ill. J. Math.} {\bf #1} (#2) #3}
\def\APP#1#2#3{{\sl Acta Phys. Polon.} {\bf #1} (#2) #3}
\def\TMP#1#2#3{{\sl Theor. Mat. Phys.} {\bf #1} (#2) #3}
\def\JPA#1#2#3{{\sl J. Physics} {\bf A#1} (#2) #3}
\def\JSM#1#2#3{{\sl J. Soviet Math.} {\bf #1} (#2) #3}
\def\MPLA#1#2#3{{\sl Mod. Phys. Lett.} {\bf A#1} (#2) #3}
\def\JETP#1#2#3{{\sl Sov. Phys. JETP} {\bf #1} (#2) #3}
\def\JETPL#1#2#3{{\sl  Sov. Phys. JETP Lett.} {\bf #1} (#2) #3}
\def\PHSA#1#2#3{{\sl Physica} {\bf A#1} (#2) #3}
\def\PHSD#1#2#3{{\sl Physica} {\bf D#1} (#2) #3}
\def\PJA#1#2#3{{\sl Proc. Japan. Acad} {\bf #1A} (#2) #3}
\def\JPSJ#1#2#3{{\sl J. Phys. Soc. Japan} {\bf #1} (#2) #3}
\newcommand{\map}{\mathcal{P}}
\def\tih{\tilde{h}}


\begin{titlepage}
\vspace*{-1cm}

\vskip 3cm

\vspace{.2in}
\begin{center}
{\large\bf A Generalized False Vacuum Skyrme model}
\end{center}

\vspace{.5cm}

\begin{center}
L. A. Ferreira$^{\dagger,}$\footnote{laf@ifsc.usp.br} and L. R. Livramento$^{\dagger, }$\footnote{livramento@usp.br}

\vspace{.3 in}
\small

\par \vskip .2in \noindent
$^{\dagger}$Instituto de F\'\i sica de S\~ao Carlos; IFSC/USP;\\
Universidade de S\~ao Paulo, USP  \\ 
Caixa Postal 369, CEP 13560-970, S\~ao Carlos-SP, Brazil\\

\normalsize
\end{center}

\vspace{.5in}

\begin{abstract}

We propose a generalization of the False Vacuum Skyrme model for any simple compact Lie groups $G$ that leads to Hermitian symmetric spaces. The Skyrme field that in the original theory takes its values in $SU(2)$ Lie group, now takes its values in $G$, while the remaining fields correspond to the entries of a symmetric, positive, and invertible $\dim G \times \dim G$-dimensional matrix $h$. This model is also an extension of the generalized BPS Skyrme model \cite{Ferreira:2025zff}. We prove that the global minima correspond to the $h$ fields being self-dual solutions of the generalized BPS Skyrme model, together with a particular field configuration for the Skyrme field that leads to a spherically symmetric topological charge density. As in the case of the original model, the minimization of the energy leads to the so-called reduced problem, defined in the context of false vacuum decay \cite{coleman1,coleman2, colemannew}. This imposes a condition on the Skyrme field, which, if satisfied, makes the total energy of the global minima, as well as the main properties of the model, equivalent to those obtained in \cite{luiz:false} for the $G=SU(2)$ case. We study this condition and its consequences within the generalized rational map ansatz and show how it can be satisfied for $G=SU(p+q)$, where $p$ and $q$ are positive integers, with the Hermitian symmetric spaces being $SU(p+q)/SU(p) \otimes SU(q) \otimes U(1)$. In such a case, the model is completely equivalent to the $G=SU(2)$ False Vacuum Skyrme model, independent of the values of $p$ and $q$. We also provide a numerical study of the baryon density, RMS radius, and binding energy per nucleon that deepens the analysis conducted in \cite{luiz:false} for the $SU(2)$ False Vacuum Skyrme model. Additionaly, in the case of $G = SU(3)$, we have studied the application of our model to the description of the binding energies and masses of the $\Lambda$-hypernuclei.

\end{abstract} 
\end{titlepage}

\section{Introduction}
\label{sec:intro}
\setcounter{equation}{0}

In the late 1950s and early 1960s, T. Skyrme proposed that baryons could be interpreted as topological solitons, with the baryon number given by the topological charge $Q$. This idea is realized in the Skyrme model, a non-linear classical field theory in the low-energy regime. In its standard formulation, the model is defined in terms of a triplet of scalar pion fields, which are assembled into the so-called $SU(2)$ Skyrme field. The baryons emerge as collective excitations of the triplet of pion fields rather than as fundamental particles.

Skyrme's innovative idea faced major obstacles due to difficulties in accurately describing the binding energies, with the model predicting binding energies roughly an order of magnitude higher than their experimental values. However, in the 1980s and 1990s, it was shown that in the large $N$-limit of QCD (Quantum Chromodynamics), where $N$ refers to the number of colors in the theory, the quark-gluon interactions become simplified, and QCD can be approximated by a theory of mesons. This boosted the interest in Skyrme idea and led to the proposal of a vast quantity of modified versions of the Skyrme model, including self-dual extensions \cite{laf2017, luiz1, quasi-self-dual, luiz:false} and $U(1)$ gauged versions that allow the introduction of electric charge by gauging a $U(1)$ subgroup of the $SU(2)$ global symmetry \cite{Callan:1983nx, Piette:1997ny, Radu:2005jp, Livramento:2023keg, Livramento:2023tmm}.

Some modifications of the Skyrme-type model, conjectured using the idea of self-duality presented in \cite{genbps}, drastically reduce the binding energies to zero. Self-duality plays an important role in the construction of topological solutions in many theories that have two key ingredients. First, the solution of the model must be classified by a topological charge $Q$ that possesses an integral form, and its density ${\cal Q}$ must have the form of a contraction ${\cal Q}\propto A_\alpha\,\tilde{A}_\alpha$ of two objects $A_\alpha$ and $\tilde{A}_\alpha$ that depend on the fields and their first-order space-time derivatives only. The meaning of the $\alpha$ index depends on the nature of fields of each theory.  Second, the static energy density ${\cal E}$ of the model must have the form ${\cal E} \propto \(A_\alpha^2 + \tilde{A}_\alpha^2\)$.

The energies of the topological solutions in self-dual theories are bounded from below by the modulus of the topological charge, up to some constant, the so-called BPS bound \cite{genbps}. On one hand, the solutions of the first-order differential equations $A_\alpha = \pm \tilde{A}_\alpha$, called self-duality or BPS equations, imply the second-order differential Euler-Lagrange equations. On the other hand, the self-dual solutions saturate the BPS bound, leading to the global energy minimizer for each value of $Q$. The self-sector formed by the solutions of the self-duality equations is a subset of the static sector composed of the static solutions of the model.

In terms of the accurate description of binding energies and nuclear radii, nowadays the best Skyrme-type model is the False Vacuum Skyrme model \cite{luiz:false}, at least for heavy nuclei. This theory is an extension of the static version of the BPS Skyrme model \cite{laf2017, luiz1}, also called the self-dual Skyrme model. Both theories are defined in terms of the Skyrme field $U$, which maps the three-dimensional space to $SU(2)$, and six scalar fields corresponding to the entries of a symmetric, invertible, and positive $3 \times 3$ matrix $h$.

The BPS Skyrme model is constructed from the topological charge using ideas of self-duality. Although the standard Skyrme model does not have self-dual solutions, as demonstrated in \cite{mantonruback, derek}, the introduction of the $h$ fields leads to an attainable BPS bound for each value of $Q$. In fact, this model possesses an infinite number of exact solutions for each value of $Q$.

The extensions of BPS models may preserve some of the self-duality equations, as it happens for the False Vacuum Skyrme model \cite{luiz:false} and the quasi-self-dual extensions of the BPS Skyrme model proposed in \cite{quasi-self-dual}. These quasi-self-dual models are particularly useful for constructing topological solitons with very low binding energies relative to the total energy, especially when the energy of the extension terms is much lower than that of the BPS terms. In nature, this two-energy scale regime is observed, for example, in nuclei, where the binding energies are of the order of $1\%$ of their total mass.  

The BPS Skyrme model is a particular case of the generalized BPS Skyrme model \cite{Ferreira:2025zff}, where the Skyrme field $U$ maps the physical space to any simple compact Lie group $G$ that leads to the Hermitian symmetric space $G/H \otimes U(1)$, where the little group $H \otimes U(1)$ is a subgroup of $G$. The simple compact Lie groups that lead to a Hermitian symmetric space are  $G=A_r,\, B_r,\, C_r,\,D_r,\,E_6,\,E_7$. In addition, $h$  becomes a $\dim  G\times \dim G$ dimensional symmetric, invertible and positive matrix. 

The generalized BPS Skyrme model is conformally invariant in three spatial dimensions and possesses some remarkable properties. For $G = SU(2)$, the six static Euler-Lagrange equations for the $h$ fields alone are equivalent to the nine self-dual equations, as proven in \cite{luiz1}.  In this work, we denote such a property as $h$-equivalence. In this case, the self-duality equations are reduced to a set of algebraically equations that can fully determine the $h$ fields in terms of the Skyrme field, which is totally free.  The freedom of the Skyrme fields remains to any other group listed above, but the $h$ matrix can no longer be fully determined by the self duality equations. In fact, the freedom of the $h$ fields tends to grows with the dimension of $G$. 

In this paper we propose generalize the $G=SU(2)$ False Vacuum Skyrme model to any simple compact Lie group $G$ that leads to the Hermitian symmetric space $G/H \otimes U(1)$, corresponding to an extension of the generalized BPS Skyrme model. The model is defined as the sum of two energy functionals $E=E_1+E_2$. The first term $E_1$ corresponds to the static energy of the generalized BPS Skyrme model. The second $E_2$ contains kinetic and potential terms to a power fraction of the topological charge density  ${\cal Q}$, plus a topological term that reproduces approximately the Coulomb interaction. Therefore, just the term $E_1$ depends on the $h$ fields. This theory acts as a classical field theory for the liquid drop model.

The $h$-equivalence of the $G = SU(2)$ BPS Skyrme model implies that the extension term $E_2$ does not break the self-dual equations in the sense that any static solution is a particular case of the self-dual solutions of \rf{self}. Consequently, for the case $G = SU(2)$, the global energy minimizer of \rf{false0} is a particular self-dual configuration. This is the main reason, used in \cite{luiz:false}, for the dynamics of the system to reduce to Coleman's false vacuum problem \cite{coleman1, coleman2}.

Although the $h$-equivalence was demonstrated only for $G=SU(2)$, in this work we prove that the global minima of the total energy $E$ can be attained by a particular type of self-dual configuration for any simple compact Lie group $G$ that leads to the Hermitian symmetric space $G/H \,\otimes\, U(1)$.  In this particular field configuration, the $h$-fields are solutions of the self-duality equations and the Skyrme field corresponds to the global minima of the energy functional $E_2$ alone.  

The term $E_1$ becomes fully determined in terms of the topological charge associated to the Skyrme field. The Euler-Lagrange equations splits in two parts. First, the variation of $E_1$ w.r.t. the $h$ and $U$ fields vanishes due the self-duality. Then the Euler-Lagrange equations associated with the $h$ fields are satisfied, since $E_2$ does not depends on the $h$ fields. Second, the variation of the term $E_2$ w.r.t. the $U$ field gives an equations that fixes the Skyrme field. 

Our model allows a two-energy scale regime, where $E_1$ is dominant in terms of the total mass of the nuclei and grows with $\mid Q \mid$. Consequently, $E_1$ gives a vanishing contribution to the binding energy per nucleon $E_B=E_{\mid Q \mid =1}-E_{\mid Q \mid }/\mid Q \mid $. Therefore, only the term $E_2$, which is a functional of the baryon density ${\cal Q}$ and its spatial derivatives, up to the topological term, contributes to the binding energy. 

The baryon density is interpreted as the density of a self-interacting fluid. The shape of this fluid of density ${\cal Q}$ is determined by the Euler-Lagrange equations w.r.t. the field $U$, which also reduces to Coleman's false vacuum problem \cite{coleman1, coleman2}. This reduction implies that ${\cal Q}$ must be spherically symmetric, which imposes an extra condition over the Skyrme field. We study how to satisfy this condition using the generalized rational map ansatz for the Skyrme field, where ${\cal Q}$ becomes the product of a radial and an angular function $F$ that must be constant. If this condition is satisfied, the energy functional is equivalent to the case $G = SU(2)$, up to a rescaling of certain coupling constants in the case $F \neq 1$.

Inside our ansatz, we can explicitly construct a special type of Skyrme field for $G=SU(p+q)$ that leads to a spherically symmetric topological charge density for any positive integer value of $p$ and $q$. Notably, our ansatz implies that $F = 1$ independently of the values of $p$ and $q$. Consequently, the topological charge density and the static energy take exactly the same form and properties as those obtained in \cite{luiz:false} for the case $G = SU(2)$. This opens the spectrum of applications of our model for $N \equiv p + q \geq 3$ in regimes where the energies depend approximately only on the particle number, interpreted as the topological charge. In addition, all the main numerical results obtained for the $SU(2)$ False Vacuum Skyrme model in \cite{luiz:false}, such as the binding energy and the nuclear radii, can be directly extended to the $SU(N)$ False Vacuum Skyrme model, with $N = p + q$, using the same values of the coupling constants. Although for each application the values of the coupling constants of the model may vary, this constitutes examples of particular solutions from $Q = 1$ to $Q = 240$ for each value of $N \geq 2$.

In the case of $G=SU(3)$, the three-flavor symmetry allows the construction of bound states that not only include the nucleons but also include the hyperons, which are baryons containing one or more strange quarks \cite{Gal:2016boi}. The hyperon family consists of the baryons $\Lambda$, $\Sigma$, $\Xi$, and $\Omega$, which contain one, one, two, and three strange quarks, respectively, with the $\Lambda$ being the lightest member. The study of hypernuclei plays a key role in understanding hyperon--nucleon (YN) interactions, thereby extending our understanding of the strong force beyond the conventional nucleon--nucleon (NN) sector, and has the potential to reveal the macroscopic composition and structure of dense astrophysical systems, most notably in neutron stars \cite{Glendenning:1984jr, Glendenning:1991es, Ozel:2016oaf, Tolos:2020aln, Burgio:2021vgk}. 

In this paper we also study the application of the Generalized False Vacuum Skyrme model to the description of the binding energies and masses of the $\Lambda$-hypernuclei. However, due to the loss of isospin symmetry, at the classical level such an approach can be reliably applied only to very heavy $\Lambda$-hypernuclei, where the mass per baryon varies slowly with the baryon number $A$.

The paper is organized as follows. In Section \ref{sec:bps}, we present an overview of the generalized BPS Skyrme model \cite{Ferreira:2025zff}. In Section \ref{sec:false}, we introduce our generalized False Vacuum Skyrme model. In Section \ref{sec:euler}, we construct solutions to the Euler-Lagrange equations corresponding to the global minima of the total energy. In Section \ref{sec:Spq}, we also study the case $G = SU(p+q)$, which leads to the Hermitian symmetric space $SU(p+q)/SU(p) \otimes SU(q) \otimes U(1)$, and construct a spherically symmetric topological charge density where $F = 1$. In Section \ref{sec:fixing}, we show how the coupling constants and integration constants can be fixed using experimental data, similarly which is done in \cite{luiz:false}. We reexamine the $G=SU(2)$ case in Section \ref{sec:su(2)app}, focusing now on identifying the choice of potential that provides the best description of the experimental values of the baryon density in the core of nuclei, significantly extending the results obtained in \cite{luiz:false}. In Section \ref{sec:su(3)app}, we study the application of our model to the description of the masses and binding energies of $\Lambda$-hypernuclei. In Section \ref{sec:conclusion}, we present our final considerations.

\section{The generalized BPS Skyrme model}
\label{sec:bps}
\setcounter{equation}{0}

Let us consider a compact simple Lie group $G$ that leads to an Hermitian symmetric space, and denote ${\cal G}$ as its Lie algebra. The Hermitian symmetric spaces are characterized by an invariant subalgebra $U(1)$ which is generated by a generator $\Lambda$. This generator can be used to construct the involutive 
\be \sigma\(T\)\equiv e^{i\,\pi\,\Lambda}\, T\, e^{-i\,\pi\,\Lambda}\;;\qquad \qquad \qquad \qquad \mbox{\rm for any}\;T\in {\cal G}
\lab{sigmadef}
\ee
which splits the Lie algebra ${\cal G}={\cal K} + {\cal P}$ into the even and odd subalgebras ${\cal K}$ and ${\cal P}$, respectively, under the involutive automorphism \rf{sigmadef}. The Hermitian nature of such symmetric spaces lies in the fact that ${\cal P}$ is even-dimensional and is further split into two parts by the generator $\Lambda$ through
\be
{\cal P}= {\cal P}_+ +  {\cal P}_-\;;\qquad\qquad \sbr{\Lambda}{E_{\pm \alpha_\kappa}}=\pm E_{\pm \alpha_\kappa}\;;\qquad\qquad E_{\pm \alpha_\kappa}\in {\cal P}_{\pm}
\lab{hermitianniceprop}
\ee
The odd subgroups ${\cal P}_{\pm}$ are generated by the $E_{\pm \alpha_{\kappa}}$ generators, where $\kappa=1,\,2,\,...,\, \frac{{\rm dim}\;{\cal P}}{2}$. The even subgroup also splits into two parts ${\cal K}={\cal H}\oplus {\Lambda}$, where $\Lambda$ generates a $U(1)_{\Lambda}$ invariant subalgebra of ${\cal K}$. Therefore, we obtain the irreducible compact Hermitian symmetric space $G/K$ with $K=H\otimes U(1)_{\Lambda}$. This leads to the orthogonality relations 
\be
 {\rm Tr}\({\cal P}\,{\cal K}\)= {\rm Tr}\({\cal P}_{+}\,{\cal P}_{+}\)={\rm Tr}\({\cal P}_{-}\,{\cal P}_{-}\)= {\rm Tr}\(\Lambda\, {\cal H}\)=0 \lab{ortho1}
 \ee
More details are given in \cite {Ferreira:2024ivq} and \cite{Ferreira:2025zff}.

The generalized BPS Skyrme model introduced in \cite{Ferreira:2025zff} is defined in $(3+1)$-dimensional Minkowski space by the action 
\be S_{BPS}=\int d^4x\left[\frac{m_0^2}{2}h_{ab}R_\mu^a R^{b,\mu}-\frac{1}{4\,e_0^2}h_{ab}^{-1}H_{\mu\nu}^aH^{b,\mu\nu}\right]\lab{skyrmebps} \ee
where the components of the Maurer-Cartan form $R_\mu$ and the quantity $H_{\mu\nu}$ correspond to
\be R_\mu^a=i\,\trace\(\pa_\mu U \,U^\dagger \,T_a\) ;\qquad H_{\mu\nu}^a=f_{abc}\,R_\mu^b\,R_\mu^c \ee 
with the generators $T_a$, $a=1,\,...,\,\dim G$, of the Lie algebra satisfying $\left[T_a,\,T_b\right]=i\,f_{abc}\,T_c$, where $f_{abc}$ is the structure constant. In addition, $h$  is a $\dim G\times \dim G$ dimensional symmetric, invertible and positive matrix, so containing $\dim G\,\(\dim G+1\)/2$ extra scalar fields corresponding to its entries.

The generators are written in the orthogonal basis, i.e., ${\rm Tr}\(T_a\,T_b\)=\kappa\, \delta_{ab}$, where the normalized trace defined by  $\trace\(T_a\,T_b\) \equiv \frac{1}{\kappa}\,{\rm Tr}\(T_a\,T_b\) = \delta_{ab}$ eliminates the dependence on the factor $\kappa$, which depends on the representation of $G$. Furthermore, in natural units, the coupling constant $m_0$ has mass dimension, while $e_0$ is dimensionless.

The finite energy solutions of \rf{skyrmebps} require that the Skyrme field must tends to a constant matrix at infinity, leading to the compactification of its domain to a three-sphere. The homotopy group of the mapping of the $3$-sphere into a simple compact Lie group $G$ corresponds to $\pi_3(G) = \IZ$. The topological charge associated with this map admits the integral representation
\be
 Q= \frac{i}{48\,\pi^2}\int d^3x\; \ve_{ijk}\,\trace\(R_i\,R_j\,R_k\)
 \lab{topcharge}
 \ee
The original BPS Skyrme model proposed in \cite{laf2017} is recovered from \rf{skyrmebps} by choosing $G=SU(2)$. By also fixing $h = \one$ in all physical space, the model reduces to the standard $SU(2)$ Skyrme model, with \rf{topcharge} giving the number of nucleons. The self-dual equations of \rf{skyrmebps} and the topological charge \rf{topcharge} inside the self-dual sector are given, respectively, by
\be
\lambda\, \tau_{cb}\,h_{ba}=\sigma_{ca}\;;\qquad\qquad Q=\frac{1}{48\,\pi^2}\int d^3x\;\sigma_{aa} \lab{self}
\ee
where 
\be \lambda = \eta \,m_0\,e_0 ;\; \qquad \sigma_{ab} \equiv  -\frac{1}{2}\,\varepsilon_{ijk}\, R_i^a\,H^b_{jk} ;\; \qquad \tau_{ab}=R_i^a\,R_i^b \lab{lambda}\ee
The sign function $\eta = \pm 1$ characterizes the (anti-)self-dual sector for $\eta=1$ ($\eta=-1)$. The BPS energy bound for the static energy $E_1$ of the model \rf{skyrmebps} is given by 
\be E_1(Q)\geq 48\,\pi^2\,\frac{\mid m_0 \mid}{\mid e_0\mid } \mid Q \mid \lab{bpsbound}\ee 
which is saturated by the solutions of the self-dual equations \rf{self}. Thus, the binding energy per nucleon, defined as $E_B \equiv E_{\mid Q\mid =1} - E_Q / \mid Q \mid$, is zero for all integer values of $Q$.

The self-dual sector of the model \rf{skyrmebps} was studied using the generalized rational map ansatz proposed in \cite{Ferreira:2024ivq}. Even within this ansatz, the self-dual equations exactly determine only some components of the $h$ matrix, while the Skyrme field remains totally free. Therefore, the static version of the Euler-Lagrange equations alone are not enough for fixing the $h$ matrix entirely neither fixing any component of the Skyrme field without the the imposition of constraints.

The arbitrariness of the Skyrme field in the generalized BPS Skyrme model leads to an infinite number of exact topological solutions for any value of $Q$.  Only for the $G = SU(2)$ BPS Skyrme model, where the Lie subgroup $H = \emptyset$, the $h$ matrix can be fully determined in terms of the Skyrme field. In fact, for $G = SU(2)$, without using any ansatz, it is possible to show that in any domain where $\tau$ is invertible, we have
\be h= \mid m_0\,e_0\mid^{-1}\,\sqrt{\det \tau}\,\tau^{-1}\ \lab{hfields}\ee
In any domain where $\tau$ is not invertible, the freedom of the system grows beyond the freedom of the Skyrme field, with some of the components of the $h$ matrix also being free, even for $G=SU(2)$.

The three-dimensional Euclidean space can be parametrized with two-spheres $(S^2)$ centered at the origin, for each value of a radial coordinate $r$. We stereographically project such spheres onto a plane with the spatial infinity identified as a point, which is parametrized by two complex coordinates $z$ and ${\bar z}$. The coordinate system $\(r\,,\,z\,,\,{\bar z}\)$ can be defined by ($z=z_1+iz_2$)
\be
x_1= r\, \frac{i\(\bar{z}-z\)}{1+\mid z\mid^2} \; ; \qquad 
x_2= r\, \frac{z+\bar{z}}{1+\mid z\mid^2} \; ; \qquad 
x_3= r\, \frac{\(-1+\mid z\mid^2\)}{1+\mid z\mid^2} \lab{coordinates}
\ee
and its metric is given by 
\be
ds^2=dr^2+ \frac{4\,r^2}{\(1+\mid z\mid^2\)^2} \, dz\,d{\bar z} \lab{metric}
\ee

Let us consider the generalized rational map ansatz proposed in \cite{Ferreira:2024ivq} for the Skyrme field given by 
\be
U= e^{i\,f\(r\)\,g\,\Lambda\,g^{-1}} \;;\qquad\qquad \qquad g\,\Lambda\,g^{-1}=\Lambda -\frac{1}{\(1+\omega\)}\left(\sbr{S}{S^{\dagger}}+i\(S-S^{\dagger}\)\right) \lab{holog}
\ee
where $\Lambda$ is defined in \rf{sigmadef}, $f\(r\)$ is a radial profile function, and $g=g\(z\,,\,{\bar z}\)$ is  an element of the compact Lie group $G$. The matrices $S$ and $S^\dagger$ that appears in \rf{holog} are either holomorphic and anti-holomorphic, respectively, i.e., $S=S(z)$ and $S^\dagger(\bar{z})$, or vice versa, and have the form 
\be S = \sum_{\kappa} w_{\kappa}\(\chi\)\,E_{\alpha_{\kappa}}  \; \qquad {\rm and} \qquad S^\dagger =\sum_{\kappa} \bar{w}_{\kappa}\(\bar{\chi}\)\,E_{-\alpha_{\kappa}}\qquad {\rm with}\qquad\chi=z,\,\bar{z} \lab{holomorphics}\ee
where $\chi=z$ ($\chi=\bar{z}$) corresponds to a (anti-)holomorphic matrix $S$. In addition, $w_{\kappa}$ and $\bar{w}_{\kappa}$ are holomorphic complex functions of the fields parametrizing the hermitian symmetric spaces $G/H\otimes U(1)_{\Lambda}$, and $E_{\alpha_\kappa}$ and $E_{-\alpha_\kappa}$ are the generators of the subalgebras ${\cal P}_+$ and ${\cal P}_-$, respectively. The ansatz \rf{holog} works for representations of the Lie algebra of $G$ where the $S$ matrix satisfies
\be
S^2=0\;;\qquad\qquad \qquad \qquad\(S\,S^{\dagger}\)\,S=\omega\,S
\lab{omegadef}
\ee
where the eigenvalue $\omega$ of $S\,S^{\dagger}$ with eigenvector $S$ is non-negative. The $g$ element used in \rf{holog} has the form
\be
g =  e^{i\,S}\,e^{\varphi\,\sbr{S}{S^{\dagger}}}\,e^{i\,S^{\dagger}}=  \one +\frac{1}{\vartheta}\, \left[i\(S+S^\dagger\) -\frac{1}{\vartheta+1} \,\left(S\,S^\dagger+S^\dagger\,S\right) \right]  \qquad  \qquad
 \lab{gdef2}
\ee
with $\varphi=\omega^{-1}\,\ln\sqrt{1+\omega}$ and $\vartheta\equiv \sqrt{1+\omega}$.

The projection of the Maurer-Cartan form $g^{-1}\,\pa_i g$ into the even and odd subspaces ${\cal K}$ and ${\cal P}$, respectively, can be performed through
\be
g^{-1}\,\pa_i g=P_i+K_i\qquad\qquad
P_i= \frac{1-\sigma}{2} g^{-1}\,\pa_i g\qquad\qquad
K_i= \frac{1+\sigma}{2} g^{-1}\,\pa_i g \lab{ws}
\ee
while $P_i$ split into the $\pm 1$ subspaces defined in \rf{hermitianniceprop}. Thus,
\be
P_i=P^{(+)}_i+P^{(-)}_i \;;\qquad\qquad \sbr{\Lambda}{P^{(\pm)}_i}=\pm P^{(\pm)}_i \;;\qquad\qquad \sbr{\Lambda}{K_i}=0\lab{wpm}
\ee
Within the ansatz \rf{gdef2}, together with \rf{holomorphics} and \rf{ws}, one obtains that
\br
P_\chi^{(+)} &=& i\,\frac{\(1+\vartheta\)^2}{\vartheta}\,\pa_\chi \(\frac{S}{\(1+\vartheta\)^2}\)\; ;\qquad\qquad P_{\bar{\chi}}^{(+)}=0 \nonumber\\
P_{\bar{\chi}}^{(-)} &=& i\,\frac{\(1+\vartheta\)^2}{\vartheta}\,\pa_{\bar{\chi}} \(\frac{S^\dagger}{\(1+\vartheta\)^2}\) \; ;\qquad\qquad P_\chi^{(-)}=0\lab{pir}
\er

In the particular cases where the square of the generator $\Lambda$ can be written as $\Lambda^2=c\,\Lambda + \frac{1}{4}\,\(1-c^2\)\,\one$, where $c$ is a real number, it  follows that  the generalized rational map ansatz \rf{holog} reduces to
\be 
U=  e^{if\,\frac{c+1}{2}}\,\left[\one+\(e^{-if}-1\)Z\right] \; ;\quad \qquad Z \equiv \frac{1+c}{2}\,\one -g\,\Lambda\,g^{-1}
\lab{Uexpanded}\ee
where $Z$ is a projector $\(Z=Z^2\)$.

From \rf{ws} and \rf{wpm}, the Maurer-Cartan form $R_i =i\,\pa_i U\,U^{-1}$ associated to the rational map \rf{holog} becomes
\be R_{i} = -V^{-1}\, \Sigma_i\, V;\;\quad V\equiv e^{-i\,f\,\Lambda/2}\,g^{-1} \;;\quad \Sigma_i \equiv \partial_if\,\Lambda- 2\,\sin \frac{f}{2}\(P^{(+)}_i-P^{(-)}_i\) \lab{Sigmadef} \ee
where, using \rf{pir} and the definition \rf{Sigmadef}, $\Sigma_i$ in the coordinates $(r,\,z,\,\bar{z})$ have the components
\br
\Sigma_r = f' \,\,\Lambda \,;\quad \qquad \Sigma_\chi = - 2\,\sin \frac{f}{2}\,P^{(+)}_\chi\,;\quad\qquad \Sigma_{\bar{\chi}} =  2\,\sin \frac{f}{2}\,P^{(-)}_{\bar{\chi}} \lab{sigmasr}
\er
where $P^{(+)}_\chi$ and $P^{(-)}_{\bar{\chi}}$ are given in \rf{pir}. Using the adjoint representation of $g$, i.e., $g\, T_a\, g^{-1}=T_{b}\,d_{ba}\(g\)  $ and \rf{Sigmadef}, we obtain 
\be
R_i^b= \trace\(T_b \, R_i\)=-\trace \(V\, T_b\,V^{-1}\, \Sigma_i\)=
-\trace\( T_c\, \Sigma_i\)\,d_{cb}\(V\) \lab{Rsigma}
\ee
The coordinates \rf{coordinates} with the metric \rf{metric} imply 
\be i\,\ve_{ijk}\,\frac{\pa r}{\pa x^i}\,\frac{\pa \chi}{\pa x^j}\,\frac{\pa \bar{\chi}}{\pa x^k} = \eta\,\frac{\(1+\mid z \mid^2\)^2}{2\,r^2} \;;\qquad {\rm with } \qquad \eta \equiv \left\{\begin{array}{ll} +1, & \chi = z \,\,\,\(S=S\(z\)\)\\ -1, & \chi=\bar{z}\,\,\,\(S=S\(\bar{z}\)\)\end{array} \right.
\ee
which together with \rf{pir}, \rf{Sigmadef} and \rf{sigmasr}, reduce the topological charge density \rf{psidef}, which can be written as ${\cal Q} = \frac{i}{4\,\pi^2}\,\ve_{ijk}\,\partial_i f\,\sin^2\frac{f}{2}\,\trace\(P^{(+)}_j\,P^{(-)}_k\)$,  to
\br
{\cal Q} & = & -\frac{\eta\,f'}{4\,\pi^2}\,\frac{\sin^2\frac{f}{2}}{r^2}\, F \;; \qquad {\rm with} \qquad F\equiv -\frac{1}{2}\,\(1+\mid z\mid^2\)^2\,\trace \( P_\chi^{(+)} \, P_{\bar{\chi}}^{(-)}\)\lab{densityq}
\er
Using the fact that $P_{\bar{\chi}}^{(-)}= -  \(P_\chi^{(+)} \)^\dagger$ (see \rf{pir}), one obtains that $F \geq 0$. In addition, the topological charge \rf{topcharge} becomes
\br
 Q&=& \,\left[\frac{f\(r\)-\sin f\(r\)}{2\,\pi}\right]_{r=0}^{r=\infty}\;Q_{{\rm top}} \; ;\quad\quad   Q_{{\rm top}} \equiv \frac{i\,\eta}{4\,\pi}\, \int dz\,d\bar{z}\;\trace\(P^{(+)}_\chi\,P^{(-)}_{{\bar \chi}}\) \lab{charge2}
\er

Using \rf{Sigmadef} and \rf{Rsigma}, the self-duality equations \rf{self} become
\be
\lambda\, \widetilde{\tau}_{cb}\,\widetilde{h}_{ba}=\widetilde{\sigma}_{ca}
\lab{sdconsequence2}
\ee
where
\be
{\tilde h}_{ab}\equiv d_{ac}\(V\)\, h_{cd}\, d^{-1}_{db}\(V\) 
 \; ;\quad {\widetilde \tau}_{ab}\equiv d_{ac}\(V\)\, \tau_{cd}\, d^{-1}_{db}\(V\) \; ;\quad {\widetilde \sigma}_{ab}\equiv d_{ac}\(V\)\, \sigma_{cd}\, d^{-1}_{db}\(V\) 
\lab{htildedef}
\ee
The adjoint representation of a compact simple Lie group is unitary and real, so $d$ is an orthogonal matrix, or in other words, $d^T = d^{-1}$. Consequently, ${\widetilde h}$ remains real and symmetric.

Using \rf{pir}, the matrices ${\widetilde \tau}$ and ${\widetilde \sigma}$ are entirely determined in terms of the profile function $f$ and the fields that parametrize $S$ and $S^\dagger$. Direct calculation shows that the components ${\widetilde \tau}_{{\cal H} T_b}$ and ${\widetilde \sigma}_{T_a\,{\cal H}}$ are zero for all $a=1,\,...,\,\dim G$. Clearly, the index ${\cal H}$ represents each generator of the subalgebra ${\cal H}$, and so on. Therefore, if ${\cal H} \neq \emptyset$, it follows that ${\widetilde h}_{{\cal H}{\cal H}}$ remains completely undetermined and ${\widetilde \tau}$ is not invertible. The self-duality equations \rf{sdconsequence2} fully determine the components ${\widetilde h}_{\Lambda \,T_a}$ through
\be {\tilde h}_{\Lambda \,\Lambda}\:\,=\alpha\,\eta\,{\rm Tr}\(P^{(+)}_\chi\,P^{(-)}_{\bar{\chi}}\);
\lab{hlb} \quad {\tilde h}_{\Lambda \,{\cal H}} = \alpha\,\eta\,{\rm Tr}\({\cal H} \,\sbr{P^{(+)}_\chi}{P^{(-)}_{\bar{\chi}}}\)
;\quad {\tilde h}_{\Lambda \,{\cal P}_{\pm}}=0 \ee
where $\alpha \equiv \frac{2\,\sin^2\frac{f}{2}}{\lambda\,f'\,{\rm Tr}\( \Lambda^2\)}\,\frac{\(1+\mid z\mid^2\)^2}{r^2}$. Note that we replaced the normalized trace with the usual trace, since the $\kappa$ factor in both the numerator and denominator cancels out. The remaining self-dual equations \rf{sdconsequence2} reduce to
\br
0 & = & \trace\({\cal P}_-\,P^{(+)}_\chi\)\,\widetilde{h}_{{\cal P}_-\,{\cal H}} = \trace\({\cal P}_+\,P^{(-)}_{\bar{\chi}}\)\,\widetilde{h}_{{\cal P}_+\,{\cal H}} = \trace\({\cal P}_-\,P^{(+)}_\chi\)\,\widetilde{h}_{{\cal P}_-\,{\cal P}_+}\lab{3H} \\
& & \trace\({\cal P}_-\,P^{(+)}_\chi\)\,\widetilde{h}_{{\cal P}_-\,{\cal P}_-}  = -\eta\,\lambda^{-1}\,f'\,\trace\({\cal P}_- P^{(+)}_\chi\) \lab{self1}\\
& & \trace\({\cal P}_+\,P^{(-)}_{\bar{\chi}}\)\,\widetilde{h}_{{\cal P}_+\,{\cal P}_+}  = -\eta\,\lambda^{-1}\,f'\,\trace\({\cal P}_- P^{(+)}_\chi\) \lab{self2}
\er
Note that, as it happens in \rf{sdconsequence2}, there is a contraction in the line index of the $\widetilde{h}$ in all the equations \rf{3H}-\rf{self2}.

The rational map ansatz drastically simplifies the self-dual equations \rf{sdconsequence2}. It directly leads to the determination of the components  $\widetilde{h}_{\Lambda\Lambda},\,\widetilde{h}_{\Lambda {\cal P}_\pm},\, \widetilde{h}_{\Lambda {\cal H}}$ in terms of the fields that parameterize the Skyrme field $U$, and reduces the other self-duality equations to the five systems of linear equations \rf{3H}-\rf{self2} for the components  $\widetilde{h}_{{\cal H} {\cal P}_\pm}, \widetilde{h}_{{\cal P}_\pm{\cal P}_\pm}$, and $\widetilde{h}_{{\cal P}_+{\cal P}_-}$. These systems of coupled linear equations for the components of the $\widetilde{h}$ matrix are not coupled with each other. Thus, there are at least a number of $2\,{\rm dim}\,{\cal P}_+\,\({\rm dim}\,{\cal P}_+-1\)$ components of $\widetilde{h}_{{\cal P}{\cal P}}$ and $2\,{\rm dim}\,{\cal H}\,({\rm dim}\,{\cal P}_+-1)$ components of  $\widetilde{h}_{{\cal H}{\cal P}}$ completely free. Consequently, only for $G=SU(2)$, where $\dim {\cal H}=0$, can the $h$ matrix be fully determined by the self-duality equations \rf{sdconsequence2}, taking the form \rf{hfields} in any domain where $\tau$ is invertible. This case was studied in \cite{luiz1}.

\section{The generalized False Vacuum Skyrme model}
\label{sec:false}
\setcounter{equation}{0}

In this paper, we propose a generalized version of the $SU(2)$ False Vacuum Skyrme model introduced in \cite{luiz:false} for any compact simple Lie group $G$ such that $G/H\otimes U(1)$ is an Hermitian symmetric space, where the little group $H \otimes U(1)$ is a subgroup of $G$. The model is defined by the static energy functional in three spatial dimensions  
\be E= E_1+E_2;\qquad\qquad E_1=\int d^3x\left[\frac{m_0^2}{2}h_{ab}R_i^a R^{b}_i+\frac{1}{4\,e_0^2}h_{ab}^{-1}H_{ij}^a\, H^{b}_{ij}\right]\lab{false0} \,\ee
where $E_1$ is the static energy of the BPS Skyrme model \rf{skyrmebps}, and the extension term $E_2$ is given by
\br
E_2=\int d^3x\;\left[\frac{\mu_0^2}{2}\,\(\partial_i\psi\)^2+V(\psi) + G_U\(U\)\,\psi^s\right] 
\lab{e2i}
\er
where $V\(\psi\)\geq 0$ everywhere, $\mu_0$ is a mass-dimensional coupling constant, and $G_U(U)$ is a non-negative functional of the Skyrme fields $U$, but not of their derivatives. The Skyrme field $U$ maps the physical space to any simple compact Lie group $G$ that leads to the Hermitian symmetric space $G/H \otimes U(1)$. Each entry of the $\dim {  G}\times \dim { G}$ dimensional symmetric, invertible and positive matrix $h$ corresponds to an scalar field. In addition,
\be \psi^s \equiv -\frac{i}{12\,\lambda^3}\,\ve_{ijk}\,{\widehat{\rm Tr}}\(R_i\,R_j\,R_k\) = - \frac{4\,\pi^2}{\lambda^3}\,{\cal Q} \lab{psidef} \ee 
is proportional to the topological charge density ${\cal Q}$ associated with \rf{topcharge}, and $2<s<6$. We shall consider the field configurations where the field $\psi$ is non-negative in all the three-dimensional Euclidean space. 

Clearly, the model \rf{false0} corresponds to an extension of the static version of the generalized BPS Skyrme model \rf{skyrmebps}, and preserves the static Euler-Lagrange equations associated with the $h$ fields. The $SU(2)$ False Vacuum Skyrme model introduced in \cite{luiz:false} is recovered by choosing $G = SU(2)$ in \rf{false0}. 

Now, we can show that the global energy minimizer of the generalized False Vacuum Skyrme model \rf{false0} corresponds to a field configuration where the $h$ fields solve the self-dual equations \rf{self} associated to the term $E_1$, while the Skyrme field corresponds to the minimizer of $E_2$. 

Despite the energy $E$ depending on $U$ and its first and second-order spatial derivatives, let us denote $E = E[U,\, h]$ to indicate the total energy associated with any field configuration $h$ and $U$, and so on. Additionally, let us denote $U_m$ and $h_m$ as the global minima of the total energy \rf{false0}, and $U_2$ the Skyrme field corresponding to the global minimum of the term $E_2$ alone. 

The key ingredient of our proof is that the self-duality equations \rf{self} can fix only some of the components of the $h$ matrix, while $U$ is completely free. Therefore, for any $h$ matrix configuration satisfying the self-duality equations \rf{self}, which we denote by $h_0$, we obtain the same total energy of the generalized False Vacuum Skyrme model \rf{false0}, independently of the Skyrme field configuration $U$.  In fact, the BPS energy bound \rf{bpsbound} is saturated, leading to
\be E_1\left[U,\,h_0\right] = E_1(Q) =  48\,\pi^2\,\frac{\mid m_0 \mid}{\mid e_0\mid } \mid Q \mid \lab{e1}\ee 

The BPS bound \rf{bpsbound} implies that $E\left[U,\,h\right] \geq E\left[U,\,h_0\right] $. Therefore, for any field configuration $h$ and $U$, it follows that 
\be E\left[U,\,h\right]  \equiv E_1\left[U,\,h\right] + E_{2}\left[ U\right] \geq E_1\left[U,\,h_0\right] + E_{2}\left[ U\right]\geq E_1\left[U,\,h_0\right] + E_{2}\left[ U_2\right] = E\left[U_2,\,h_0\right] \lab{ine}\ee
Choosing the arbitrary field in \rf{ine} as $U=U_m$ and $h=h_m$, it follows that $E\left[U_m,\,h_m\right]  \geq E\left[U_2,\,h_0\right]$. However, by definition, $U_m$ and $h_m$ correspond to the global minimizer of the energy \rf{false0}, so $E\left[U_m,\,h_m\right]  \leq E\left[U_2,\,h_0\right]$. Therefore, we must have $E\left[U_m,\,h_m\right]  =  E\left[U_2,\,h_0\right]$. Consequently, the minima of the total energy \rf{false0} can be attained by any self-dual field configuration $h_0$ together with the Skyrme field configuration $U_2$, completing the proof. 

\section{The Euler-Lagrange equations}
\label{sec:euler}
\setcounter{equation}{0}

The global energy minimizers of the generalized False Vacuum Skyrme model \rf{false0} satisfies the same self-duality equations of the generalized BPS Skyrme model \rf{skyrmebps}. This drastically simplifies the study of the static sector of the model \rf{false0} once we can assume the self-duality equations \rf{self}.

In \cite{Ferreira:2025zff} its is explicitly shown that the self-duality equations \rf{self} satisfies the Euler-Lagrange equations associated to the $h$ and $U$ fields in the static sector of the generalized BPS Skyrme model \rf{skyrmebps}, which can be written as $\delta_h E_1=0$ and $\delta_U E_1=0$, respectively. Note that the variation of $E$ and $E_1$, as given in \rf{false0},  w.r.t the $h$ fields coincides, i.e., $\delta_h E=\delta_h E_1$. Therefore, any solution of the self-duality equations \rf{self} satisfies automatically the Euler-Lagrange equations associated to the $h$ fields in the case of the full theory \rf{false0}.

The contribution of the term $E_1$ to the Euler-Lagrange equations associated to the Skyrme field for the full model \rf{false0} vanishes automatically due the self-duality equation, which implies $\delta_U E_1=0$. Since the self-duality equations does not fixes any component of the Skyrme field, the Euler Lagrange equations w.r.t. Skyrme field comes from the term $E_2$ alone.

The topological term of \rf{e2i} does not contribute to the Euler-Lagrange equations associated with the Skyrme field, since it satisfies by construction $\delta_U\int d^3x\;G\(U\)\,\psi^s=0$. The variation of \rf{psidef} w.r.t. $U$ is given by
\be
\delta \psi^s = \frac{i}{8\,\lambda^3}\,\pa_k\left[\varepsilon_{ijk}\,H_{ij}^a\,\(\delta U\,U^{-1}\)_a\right]
\ee
Using also $\delta \psi^s = s\, \psi^{s-1}\,\delta \psi$, the variation of $E_2$ given in \rf{e2i} leads to
\br
\delta E_2 &=& \int d^3x\;\left[-\mu_0^2\,\partial_l^2\psi+\frac{\delta V\(\psi\)}{\delta \psi}\right]\,\delta \psi\nonumber\\
&=& \int d^3x\;\pa_k\left[\frac{1}{s\,\psi^{s-1}}\,\(-\mu_0^2\,\partial_l^2\psi+\frac{\delta V\(\psi\)}{\delta \psi}\)\right]\,\frac{1}{8\,\lambda^3}\,\varepsilon_{ijk}\,H_{ij}^a\,\(i\,\delta U\,U^{-1}\)_a \qquad
\lab{e2variation}
\er
where we neglect the surface term. Therefore, we obtain $\delta E_2 =0$  if the term in the brackets on the r.h.s. of the last line of \rf{e2variation} is an arbitrary integration constant $c$. The Euler-Lagrange equations associated with the Skyrme field becomes
\br
\mu_0^2\,\partial^2 \psi -\frac{\delta\,V_{\rm eff.}}{\delta\,\psi}=0\;;\qquad\qquad V_{\rm eff.}\equiv V-c\, \psi^{s}
\lab{psieq}
\er

Clearly, if we take $\psi$ as the fundamental degree of freedom instead of the $U$ field, the equation \rf{psieq} also corresponds to the Euler-Lagrange equation associated with the $\psi$ field obtained from the effective energy functional 
\br
E_{\rm eff.}= T + U_{\rm eff.}\;; \qquad\qquad T\equiv \frac{\mu_0^2}{2}\,\int d^3x\,\(\partial_i\psi\)^2 \;; \qquad\qquad U_{\rm eff.}\equiv \int d^3x\, V_{\rm eff.}
\lab{eeff}
\er
Under scale transformation $\psi_{\alpha}\(x\)=\psi\(x/\alpha\)$, we have that $T\rightarrow\alpha\,T$ and $U_{\rm eff.}\rightarrow\alpha^3\,U_{\rm eff.}$.  The stable solutions of \rf{psieq} under  Derrick's argument \cite{derrick,coleman2} satisfy 
\be 
T+3\,U_{\rm eff.}=0 ;\qquad\qquad{\rm and\:\, so}\qquad\qquad E_{\rm eff.}=\frac{2}{3}\,T
\lab{derrick}
\ee 
Since $\psi^s$ is proportional to the topological charge density \rf{psidef}, it follows that for non-vanishing topological charge configurations,  it cannot be constant over its entire domain $\IR^3$. Therefore, due to the definition \rf{eeff}, we have $T>0$. It so follows from \rf{derrick} that $E_{\rm eff.} >0$ and $U_{\rm eff.}<0$. Consequently, $V_{\rm eff.}$ must be negative somewhere, and using the fact that the physical potential $V$ is non-negative, we obtain that $c$ must be positive. In fact, depending of the definition of the original potential $V(\psi)$, there may be a critical positive value for the integration constant, denoted by $c_{\rm crit.}$, such that for $V_{\rm eff.} $ to be negative somewhere, we must have $c > c_{\rm crit.}$.

From now on, we consider the choice of the physical potential $V$ and the integration constant $c$ such that $V_{\rm eff.}$ is an {\it admissible }potential \cite{coleman2}, i.e.: (i) $V_{\rm eff.}$ is continuously differentiable for all $\psi$; (ii) $V_{\rm eff.}\(0\)= \pa_\psi V_{\rm eff.}\(0\)= 0$; (iii) $V_{\rm eff.}$ is somewhere negative; (iv) there exist positive numbers $a_1,\, a_2,\, b_1,\, b_2$ such that $b_1< b_2 < 6$ and $V_{\rm eff.} -a_1\, \mid \psi \mid^{b_1}+a_2\, \mid \psi \mid^{b_2} \geq 0$ for all values of $\psi$.  In addition, we must assume that $V$  is non-negative everywhere and has a non-degenerate vacuum at $\psi=0$, which due the conditions above corresponds to a local maximum of $V_{\rm eff.}$. Thus, for $E_{\rm eff.}$ to be finite, we must consider the field configurations that vanish at spatial infinity, which we denote as $\psi(\infty)=0$. 

The task of obtain the minimizer of $E_{\rm eff.}$ with the  boundary condition $\psi(\infty)=0$ corresponds to the reduced problem defined in \cite{coleman2}, which was studied by S. Coleman, V. Glaser, and A. Martin, and it is of great importance in the study of the false vacuum decay \cite{coleman1, coleman2, colemannew}. {\it The main theorem } presented in \cite{coleman2} shows that the equation \rf{psieq} possesses at least one monotone spherically symmetric solution $\psi_m$ vanishing at infinity, which is called bounce solution, other than the trivial solution $\psi=0$ everywhere. The effective energy \rf{eeff} evaluated by such a solution $E_{\rm eff.}\left[\psi_m,\,\pa_r \psi_m\right]$ is less than or equal to the effective energy associated with any other solution of \rf{psieq}  vanishing at infinity.  

The effective energy \rf{eeff} and the term $E_2$ defined in \rf{e2i} differ only by the topological term $E_{{\rm top}}\equiv \int d^3x\,\(G+c\)\,\psi^s$, which is constant within a given homotopy class. Consequently, for each value of the topological charge \rf{topcharge}, the minima of $E_2$ and $E_{\rm eff.}$ coincide. Therefore, the minima of $E_2$ and the total energy \rf{false0} correspond to a monotone spherically symmetric field $\psi$ configuration. The remaining Euler-Lagrange equations \rf{psieq} reduce to an ordinary differential equation for a radial function $\psi\(r\)$, given by
\be
\mu_0^2\left[\frac{d^2\,\psi}{d\,r^2}+\frac{2}{r}\,\frac{d\,\psi}{d\,r}\right]-\frac{\delta\,V_{\rm eff.}}{\delta\,\psi}=0
\lab{psieqradial}
\ee

The field $\psi$ \rf{psidef} is directly proportional to the topological charge density ${\cal Q}$. Therefore, we must find an ansatz for the Skyrme field that leads to a spherically symmetric topological charge density. Using the generalized map ansatz \rf{holog}, the topological charge density becomes \rf{densityq}, which is spherically symmetric just when the function $F$ of the complex coordinates $z$ and $\bar{z}$ is a  constant. 

Using the generalized map ansatz \rf{holog}, the self-duality equations \rf{self} are reduced to \rf{hlb}-\rf{self2}, as shown in Section \ref{sec:bps}. The self-duality equations \rf{hlb} fixes the components ${\tilde h}_{\Lambda \,\Lambda},\, {\tilde h}_{\Lambda \,{\cal H}},\, {\tilde h}_{\Lambda \,{\cal P}_{\pm}}$. In this work, we choose to solve the remaining self-duality equations \rf{3H}-\rf{self2} by choosing
\be
\widetilde{h}_{{\cal P}_\pm\, {\cal H}} = 0;\;\qquad\qquad \widetilde{h}_{{\cal P}\, {\cal P}}  = - \eta\,\lambda^{-1}\,f'\,\one \lab{5}
\ee
where the second equation of \rf{5} is equivalent to $\widetilde{h}_{{\cal P}_\pm\, {\cal P}_\pm}  = - \eta\,\lambda^{-1}\,f'\,\one$ and $\widetilde{h}_{{\cal P}_\mp\, {\cal P}_\pm}  = 0$. Although $\widetilde{h}_{{\cal H}\, {\cal H}}$ is totally free by the self-duality equations, we can choose $\widetilde{h}_{{\cal H}\, {\cal H}}  = - \eta\,\lambda^{-1}\,f'\,\one$ to fix the $\widetilde{h}$ matrix entirely. Once the Skyrme field is determined, the matrix $V$ defined in \rf{Sigmadef} is fixed. Therefore, in case $\widetilde{h}$ is fully determined, the matrix $h$ is also fully determined through \rf{htildedef}.

From now on, we will assume that we know the $S\(\chi\)$ matrix satisfyng  \rf{omegadef} that leads, using \rf{pir} and \rf{densityq}, to 
\be
F = \frac{1}{2}\,\(1+\mid z\mid^2\)^2\,\trace \( P_\chi^{(+)}\, \(P_\chi^{(+)} \)^\dagger\)  =  {\rm const.} >0 \lab{condition}
\ee
On the other hand, to generalize the results of \cite{luiz:false}, we must consider the cases where $\mid Q_{{\rm top}} \mid = 1 $, i.e.,
\be
\mid Q_{{\rm top}} \mid = 1 \qquad \Rightarrow \qquad Q= {\rm sign} \(Q_{{\rm top}}\) \,\left[\frac{f\(r\)-\sin f\(r\)}{2\,\pi}\right]_{r=0}^{r=\infty}  \lab{charge3b}
\ee
where we use \rf{charge2}. Since we want to interpret $\mathcal{Q}$ as the nuclear matter density, we shall restrict \rf{charge3b} to the cases where $Q > 0$. Now, let us fix the boundary conditions for the profile function through
\be
\left. \begin{array}{c} f\(0\)=0 \;\;{\rm and }\;\; f\(\infty\)=2\,\pi\, M \;\; {\rm for }\;\;{\rm sign} \(Q_{{\rm top}}\)=+1 \\ f\(0\)=2\,\pi\, M \;\;{\rm and }\;\; f\(\infty\)=0 \;\; {\rm for }\;\;{\rm sign} \(Q_{{\rm top}}\)=-1 \end{array}\right\} \quad \quad\Rightarrow \quad \quad Q=M \lab{sign}
\ee

The analysis of the equation \rf{psieqradial} is straightforward since, as Coleman points out \cite{coleman1, coleman2}, it can be interpreted as the equation of motion of a mechanical particle moving under the influence of the inverted potential $V_{\rm eff.}^{(-)}\equiv -V_{\rm eff.}$, and a viscous force $-\mu_0^2\,\frac{2}{r}\,\frac{d\,\psi}{d\,r}$, where $\psi$ represents the spatial coordinate and $r$ the time. In fact, multiplying \rf{psieqradial} by $\frac{d\,\psi}{d\,r}$, one obtains that
\be
\frac{d\,{\cal E}_{\rm particle}}{d\,r}=-\mu_0^2\,\frac{2}{r}\,\(\frac{d\,\psi}{d\,r}\)^2
 \;;  \qquad\quad {\rm with}\qquad \quad {\cal E}_{\rm particle}=\frac{\mu_0^2}{2}\,\(\frac{d\,\psi}{d\,r}\)^2+V_{\rm eff.}^{(-)}\qquad \lab{eparticleder}
\ee
where ${\cal E}_{\rm particle}$ is the mechanical energy of the particle. The particle starts at a position $\psi(0)$ with zero velocity $\frac{d\,\psi}{d\,r}\mid_{r=0}=0$, and must dissipate all its initial potential energy until it reaches $\psi=0$ $(V_{\rm eff.}=0)$ at infinity time $r\rightarrow \infty$.

The nuclear matter density has an experimentally observed exponential asymptotic decay, which can be incorporated into the model \rf{false0} by choosing a physical potential $V$ that includes a mass term for the $\psi$ field. Writing $V=\beta_2^2\, \psi^2 + \widetilde{V}$, where $\beta_2$ is a coupling constant, it is sufficient to choose
\br
V_{\rm eff.}= \beta_2^2\,\psi^2-c\,\psi^s+{\widetilde V};\qquad \qquad2<s<6 
\lab{potminus1}
\er
and 
\be
{\widetilde V}\geq 0;\qquad \quad {\widetilde V}\mid_{\psi=0}=0;\qquad \quad\frac{\delta\,{\widetilde V}}{\delta\,\psi}\mid_{\psi=0}=0
;\qquad\quad \frac{\delta^2\,{\widetilde V}}{\delta^2\,\psi}\mid_{\psi=0}=0
\ee
to ensure that $V_{\rm eff.}$ is admissible potential. However, the existence of a critical value $c_{\rm crit.}$, such that $V_{\rm eff.}$ is an admissible potential if and only if $c > c_{\rm crit.}>0$, plays a key role in reproducing the experimental behavior of the nuclear matter density, as we shown bellow.  In this case, for $c = c_{\rm crit.}$, the true vacuum of the effective potential becomes degenerate with the false vacuum $\psi = 0$ in the cases where $c \leq  c_{\rm crit.}$. 

When $c$ becomes very close to $c_{\rm crit.}$ from the left, it follows that the inverted potential  $V_{\rm eff.}^{(-)}\rightarrow 0^+$. Therefore, the total mechanical energy of the particle, which vanishes for $r \rightarrow \infty$, tends to zero at $r = 0$. By integrating the first equation in \rf{eparticleder} and using these boundary conditions for the total mechanical energy, we obtain that the dissipated energy due to the viscous force must vanish. Thus, the particle must maintain its approximately zero velocity from $r = 0$ until a very large radius, when the term $1/r$ of the viscous force becomes much larger than the square of the velocity, suppressing the viscous force. Therefore, the field $\psi$, and consequently ${\cal Q}$, develop a plateau from $r = 0$ to $r = R(c)$, where $\psi$ is practically constant, and they decay quickly to the false vacuum for $r>R\(c\)$. The rms radius becomes
\be
\sqrt{\langle r\rangle}=\sqrt{\frac{\int d^3x\, r^2\,{\cal Q}}{\int d^3x\,{\cal Q}}}\sim \sqrt{\frac{3}{5}}\,R\(c\); \qquad \rightarrow \qquad \sqrt{\langle r\rangle}\propto Q^{1/3}
\lab{rqonethird}
\ee
which corresponds to the power law observed experimentally for nuclei above a certain mass number. This also agrees with the experimental behaviour of the nuclear matter density, which varies slowly, being approximately constant from its initial value at the origin of the nucleus until some radius, beyond which it falls exponentially.

To ensure the existence of the critical value, we must also choose the physical potential $V$ such that there is a positive value of $\psi$, denoted by $\psi_m$, for which $V(\psi) > 0$,  $\forall \psi > \psi_m$. All the above considerations on $V$ can be fulfilled by choosing
\br
V= \beta_2^2\,\psi^2 + \beta_{\kappa}^2\,\psi^{\kappa};\quad 6>s>2;\;\;\kappa>s \lab{pot0}
\er
where $\beta_\kappa$ is a coupling constant, and $\widetilde{V} = \beta_{\kappa}^2\,\psi^{\kappa}$. Thus, the effective potential given in \rf{potminus1} is admissible. Using the dimensionless quantities 
\be 
\zeta=\frac{\beta_2}{\mu_0}\,r;\qquad\qquad\hpsi=\(\frac{\beta_{\kappa}^2}{\beta_2^2}\)^{\frac{1}{\(\kappa-2\)}}\,\psi ;\qquad\qquad \gamma=\frac{c}{\beta_2^2} \,\(\frac{\beta_2^2}{\beta_{\kappa}^2}\)^{\frac{s-2}{\kappa-2}} \lab{dimensionless}
\ee
one obtains that \rf{psieq} becomes 
\be
\frac{d^2\,\hpsi}{d\,\zeta^2}+\frac{2}{\zeta}\,\frac{d\,\hpsi}{d\,\zeta}- \frac{\delta \widehat{V}_{{\rm eff.}}}{\delta \hpsi}=0;\qquad\qquad \widehat{V}_{{\rm eff.}}=\widehat{V}-\gamma\,\hpsi^{s}=\hpsi^2-\gamma\,\hpsi^{s}+\hpsi^{\kappa}
\lab{hpsieq}
\ee
which depends only on the rescaled integration constant $\gamma$, the fixed parameter $\kappa$ of the potential, and the proportionality power factor $s$ between $\psi$ and ${\cal Q}$ introduced in \rf{psidef}. The critical value $\gamma_{\rm crit.}$ of the integration constant, for which the effective potential $\widehat{V}_{{\rm eff.}}$ becomes non-negative everywhere and its vacuum becomes degenerate, and the vacuum $\hpsi_{\rm crit.}$ that does not correspond to the trivial one $\hpsi = 0$, are given respectively by
\be \gamma_{\rm crit.}= \(\frac{\kappa-2}{\kappa-s}\)\,\(\frac{\kappa-s}{s-2}\)^\frac{s-2}{\kappa-2};\qquad\qquad \hpsi_{\rm crit.}= \(\frac{s-2}{\kappa-s}\)^\frac{1}{\kappa-2} 
\lab{critical}
\ee
By varying $\gamma$, we can construct solutions for distinct values of the topological charge given in \rf{topcharge}, which, using \rf{hpsieq}, can be written as
\br
Q =\(\frac{\mu_0}{\beta_2}\)^3\,\vartheta\; I\(\gamma\,,\,s\,,\,\kappa\) \;;\quad \vartheta=\(m_0\,e_0\)^3\,\(\frac{\beta_2^2}{\beta_{\kappa}^2}\)^{\frac{s}{\(\kappa-2\)}} \;; \quad  I=\frac{1}{\pi}\int_0^{\infty}d\zeta\,\zeta^2\,\hpsi^s
\lab{chargefinal}
\er
and the rms radius is given by
\be
\sqrt{\langle r^2\rangle}\equiv\sqrt{\frac{\int d^3x\, r^2\,\psi^s}{\int d^3x\, \psi^s}}=\frac{\mu_0}{\beta_2}\,\Lambda\(\gamma\,,\,s\,,\,\kappa\)
\lab{radius}
 \ee 
 with 
 \be
 \Lambda \equiv \sqrt{\frac{J\(\gamma\,,\,s\,,\,\kappa\)}{I\(\gamma\,,\,s\,,\,\kappa\)}};\qquad \qquad 
 J=\frac{1}{\pi}\int_0^{\infty}d\zeta\,\zeta^4\,\hpsi^s
 \lab{LambdaJdef}
 \ee
From \rf{psidef}, \rf{dimensionless} and the definition of $\vartheta$ given in \rf{chargefinal}, the topological charge density, which is interpreted as the nuclear density, become 
\be {\cal Q} =  \frac{\vartheta}{4\,\pi^2}\,\hpsi^s \lab{nucleardensity}
\ee 

As demonstrated in \cite{luiz:false}, when the integration constant tends to its critical value from the left, i.e., $\gamma \rightarrow \gamma_{\rm crit.}^+$, the solutions of \rf{hpsieq} imply $Q \rightarrow +\infty$. Therefore, in the left neighbourhood of the critical value of $\gamma$, one can construct infinitely large nuclei. However, a key point not explored in \cite{luiz:false} is that the nuclear density of such nuclei must tend to a determined critical density as $\gamma$ approaches its critical value. In fact, when $\gamma \rightarrow \gamma_{\rm crit.}^+$, we have $\hpsi (r=0) \rightarrow \hpsi_{\rm crit.}^+$. Therefore, once the $\vartheta$ parameter is fixed using experimental nuclear data, which depends on the choice of the values of $s$ and $\kappa$, and such parameters are also fixed,  the topological charge density at the origin obtained from \rf{nucleardensity} becomes
\be {\cal Q}_{{\rm crit.}}(r=0) =  \frac{\vartheta}{4\,\pi^2}\,\hpsi^s_{{\rm crit.}} = \frac{\vartheta}{4\,\pi^2}\,\(\frac{s-2}{\kappa-s}\)^\frac{s}{\kappa-2}  \lab{criticaldensity}
\ee 

The trace of the Skyrme field within the general rational map ansatz \rf{holog} is given by ${\rm Tr}\,\(U\) =e^{i\,f(r)\,\Lambda}$ and depends only on the profile function. We shall consider $G_U\(U\)$ as a non-negative functional only of $\Tr\,\(U\)$. This breaks the global $G_L\otimes G_R$ symmetry to the diagonal $G$ subgroup. On the other hand, from \rf{psidef} and $g\equiv\(f-\sin f\)$,  where $\frac{dg}{dr}= - 2\,f'\,\sin^2\(\frac{f}{2}\)$, we can write
\br
\psi^s=-\frac{F}{\(m_0\,e_0\)^{3}}\,\frac{1}{2\,r^2}\,\frac{dg}{d\,r}\lab{psig}
\er
We take $G_U$ as a function of $g$, and the topological term of \rf{false0} can be written as
\be
Q_G\equiv\int d^3x\, G_U\,\psi^s= - \frac{2\,\pi\,F}{\(m_0\,e_0\)^{3}}\,\left[P\(g\(\infty\)\)-P\(g\(0\)\) \right];\qquad {\rm with}\qquad \frac{d\,P}{d\,g}=G_U
\ee
Now, choosing $G_U={\rm sign} \(Q_{{\rm top}}\)\,\beta_G^2\,g/\(2\,\pi\)$, which implies $P={\rm sign} \(Q_{{\rm top}}\)\,\beta_G^2\,g^2/\(4\,\pi\)+{\rm const.}$, and using \rf{sign} and \rf{dimensionless}, we obtain
\be
Q_G = \frac{\sigma_2\,\sigma_G}{2} \,Q^2 \;;\qquad {\rm where}\qquad \sigma_2=\frac{4\,\pi^2\,\beta_2^2}{m_0^3\,e_0^3}\,\(\frac{\beta_{\kappa}^2}{\beta_2^2}\)^{\frac{s-2}{\kappa-2}};   \qquad
\sigma_G=F\,\frac{\beta_G^2}{\beta_2^2}\,\(\frac{\beta_2^2}{\beta_{\kappa}^2}\)^{\frac{s-2}{\kappa-2}} \lab{couloumb}
\ee
Using the choice \rf{sign} for which $Q>0$, together with \rf{e2i}, \rf{e1}, \rf{dimensionless} and \rf{couloumb}, the total energy \rf{false0} reduces to 
\be 
E= 48\,\pi^2\,\frac{\mid m_0 \mid}{\mid e_0\mid }\,  Q  + E_2\;;\qquad \qquad E_2= \sigma_2\,\left[\vartheta\,\(\frac{\mu_0}{\beta_2}\)^3\, {\cal E}+ \frac{\sigma_G}{2} \, Q^2
\right] \lab{energyf}
\ee
where
 \be
 {\cal E}=\frac{1}{\pi}\int_{0}^{\infty} d\zeta\,\zeta^2 \, \left[\frac{\hpsi^{\prime\,2}}{2}+\hpsi^2+\hpsi^{\kappa}\right]\:
 \lab{energycal}
 \ee
The binding energy per nucleon $E_B \equiv E_{\mid Q\mid =1} - E_Q / \mid Q \mid$ becomes
\be
E_B=\sigma_2\,\left[\vartheta\(\frac{\mu_0}{\beta_2}\)^3\( {\cal E}_{Q=1}-\frac{{\cal E}}{Q}\)+\frac{\sigma_G}{2}\(1- Q\)
\right]
\lab{binding}
\ee

The effect of a positive constant $F$ in the total energy \rf{energyf} in relation to the energy of the original $SU(2)$ False Vacuum Skyrme model, where $F=1$, is just the rescaling of the some coupling constants. In fact, $F$ can be absorbed into $\beta_G^2$ in \rf{couloumb} reducing the energy \rf{energyf} to its $G=SU(2)$ form, as long we can satisfy \rf{condition}. This rescaling does not change the equations \rf{hpsieq} or \rf{chargefinal}. Therefore, the bounce solutions of \rf{hpsieq} and their corresponding topological charge \rf{chargefinal} obtained for $G = SU(2)$, can be interpreted as solutions for larger groups, provided that the condition \rf{condition} is satisfied. However, the shape of the profile function obtained by solving \rf{psidef} with the boundary conditions \rf{sign} may depend on the value of $F$.

The energy \rf{energyf} does not depend explicitly on the number of fields that parametrizes the $G$ Lie group, since it is fully determined by ${\cal Q}$ that depends only on the profile function, due to \rf{densityq} and \rf{condition}. This is a profound consequence of the construction of the $E_2$ term, defined only in terms of the topological charge density and its first-order derivatives, the self-duality equations \rf{self}, and the condition \rf{condition}. However, to satisfy this condition, we may have to fix the fields that parametrize the coset $G/K$.

Notably, if one can construct an $S$ matrix \rf{holomorphics} that satisfies \rf{omegadef} and simultaneously reduces the angular function \rf{condition} to $F = 1$, then the total energy \rf{energyf}, the Euler-Lagrange equation for the field $\psi$ given in \rf{psieqradial}, and the profile function $f(r)$, obtained from \rf{psig}, all reduce completely to their $G = SU(2)$ form. This reduction makes it possible to extend the results obtained in \cite{luiz:false} for the $SU(2)$ case, specifically the energy \rf{energyf}, binding energy per nucleon \rf{binding}, rms radius \rf{radius}, and baryonic density \rf{densityq} of the false vacuum Skyrmions, to anothers compact simple Lie groups. 

A major consequence of the $SU(2)$ reduction is that, by employing the same values of the coupling constants used in \cite{luiz:false}, which were determined from experimental data, our model can also accurately reproduce the experimental values of the binding energy and the RMS radius for larger groups, provided that the mass number satisfies $A \geq 12$. Remarkably, in Section \ref{sec:Spq} we present an explicit example of the construction of the $S$ matrix satisfying $F=1$, within the rational map ansatz \rf{holog}, for $G=SU(p+q)$ and the Hermitian symmetric space $SU(p+q)/SU(p) \otimes SU(q) \otimes U(1)$, valid for any pair of positive integers $p$ and $q$.

The key ingredient of the $SU(2)$ reduction is the fact that self-duality eliminates the explicit dependence of the $h$ fields in the energy functional \rf{false0}, reducing the model to a classical field theory for the $\psi$ field, up to some topological terms, as given in \rf{energyf}. However, this holds only for global energy minimizers, which must satisfy the self-duality equations, as we have shown above. On the other hand, the fields $h$ and the explicit form of the Skyrme field still depend on the Lie group $G$.

The reduction to the $SU(2)$ False Vacuum Skyrme model also highlights the importance of the topological term in \rf{e2i}, which can be chosen so that it gives a positive contribution that is proportional to $Q^2$ in the total energy. For heavy nuclei, where the neutron-to-proton ratio changes slowly with mass number, this term approximately reproduces the Coulomb interaction, which is the main physical reason why the binding energy per nucleon tends to decrease above the mass of $^{56}$Fe. Note that, since we are considering the baryon number as a topological quantity (topological charge),  it is also natural to treat the electric charge as a topological quantity.

The topological term in \rf{energyf} is the key reason why the binding energy per nucleon decreases approximately linearly with $Q$ for heavy nuclei in our model, although at the classical level the model \rf{false0} does not distinguish between protons and neutrons. In fact, without this topological term ($\sigma_G = 0$), the binding energy per nucleon $E_B$ increases monotonically with $Q$ and saturates as $Q \to \infty$, as shown in $\cite{luiz:false}$.

In the cases that $\sigma_G$ is small enough, $E_B$ for light nuclei is monotonically increasing. However, the Coulomb interaction gets stronger when the topological charge grows. For some value of $Q$ the binding energy per nucleon \rf{binding} gets its global maximum  and falls approximately linearly to large values of $Q$. The beauty of the model is that theses properties of the binding energy per nucleon, as well the growing of the rms radius with $Q^{1/3}$ to large values of $Q$ (see \rf{rqonethird}), are robust by change of the physical potential. This is a consequence of two facts: $\gamma(Q)$ being monotonically decreasing, which is observed numerically; the choice of the physical potential that leads to an admissible potential. 

\section{The case of $SU\(p+q\)/SU\(p\)\otimes SU\(q\)\otimes U\(1\)$} \label{sec:Spq}
\setcounter{equation}{0}

The construction of the $S$ matrix defined in \rf{holomorphics} satisfying the conditions \rf{omegadef} for the Hermitian symmetric space $SU\(p+q\)/SU\(p\)\otimes SU\(q\)\otimes U\(1\)$ was done in \cite{Ferreira:2024ivq}. It works for the fundamental $(p+q) \times (p+q)$ representation of the $G=SU(p+q)$ Lie group, where $p$ and $q$ are positive integers. The $\Lambda$ and $S$ matrices constructed in \cite{Ferreira:2024ivq} for such a case can be written as  
\be
\Lambda = \frac{1}{p+q}\,\(\begin{array}{cc} q\,\one_{p\times p} & O_{p\times q}\\ O_{q\times p} & -p\,\one_{q\times q}\end{array}\)\qquad\quad {\rm and}\quad \qquad  S = \(\begin{array}{cc} O_{p\times p} & u \, \otimes\, v\\ O_{q\times p} & O_{q\times q}\end{array} \) \lab{Spq}
\ee
where $O_{p\times q}$ corresponds to a $p\times q$ zero matrix, and so on. Additionally, the $S$ matrix given in \rf{Spq} is parametrized by $p$ complex scalar fields $u_a$, with $a=1,\,...,\,p$, and $q$  complex scalar fields $v_b$, with $b=1,\,...,\,q$, which are the entries of $u^T=\(u_1,\,...,\,u_p\)$ and $v^T=\(v_1,\,...,\,v_q\)$. The eigenvalue of $S\,S^\dagger$ corresponding to the eigenvector $S$, as defined in \rf{omegadef}, is $\omega = \mid u \mid^2 \,\mid v \mid^2$.

In order for \rf{Spq} to lead to $F = {\rm const.}$, we can take, as done in \cite{Ferreira:2024ivq}, the following choice
\br
u_c&=& u_1\(\chi\);\qquad  v_d= v_1\(\chi\); \qquad  w\equiv \sqrt{p\,q}\,u_1\,v_1 \lab{restrictive}\er 
for all $c=1,\,...,\,p$ and $d=1,\,...,\,q$,  where the $w$ field is fixed through
\br
w=e^{i\,\alpha}\,\chi \qquad\qquad {\rm or} \qquad\qquad  w=\frac{\beta\(\chi-\mid \beta \mid^{-1}\,e^{i\alpha}\)}{\chi+\mid \beta \mid\,e^{i\alpha}} \lab{restrictive2}
\er
where $\alpha$ is a real constant contained in the interval $\left[0,\,2\pi\)$, and $\beta$ is an arbitrary non zero complex constant. Within the field configuration \rf{restrictive} and \rf{restrictive2}, we obtain from \rf{pir} and \rf{Spq} that 
\br 
S & =& \frac{w}{\sqrt{p\,q}} \, \(\begin{array}{cc} O_{p\times p} & I_{p\times q}\\ O_{q\times p} & O_{q\times q}\end{array}\)\, \qquad \Rightarrow \qquad P_\chi^{(+)} =  i\,\frac{\vartheta^{-2}}{\sqrt{p\,q}} \, \pa_\chi w \,\(\begin{array}{cc} O_{p\times p} & I_{p\times q}\\ O_{q\times p} & O_{q\times q}\end{array}\) 
\er
where $I_{p\times q}$ is a $p\times q$ matrix with all entries equal to $1$. Therefore, the topological charge density and the $F$ function given in \rf{densityq} become
\br
{\cal Q} & = & -\frac{\eta\,f'}{8\,\pi^2}\,\frac{\sin^2\frac{f}{2}}{r^2} \;; \qquad {\rm with} \qquad F= \(\frac{1+\mid \chi \mid^2}{1+\mid w \mid^2}\)^2 \,\pa_\chi w\,\pa_{\bar{\chi}} \bar{w}=1\lab{densityq2}
\er

Note that we have $F = 1$ for both the $w$ fields given in \rf{restrictive2}, and the topological charge defined in \rf{charge2} for those maps $w$ between two-spheres corresponds to $Q_{{\rm top}} \equiv \eta \, {\rm deg}\,w= \eta$. Therefore, not only is the condition \rf{condition} satisfied, but we also have the same value of $F$ as in the original $SU(2)$ case, where $F=1$. As a result, there is a total reduction to the $SU(2)$ False Vacuum Skyrme model, even without rescaling the coupling constant $\beta_G$ to absorb $F$. Therefore, as explain in Section \ref{sec:euler}, there is a total reduction to the $SU(2)$ case studied in \cite{luiz:false} in terms binding energy per nucleon \rf{binding}, rms radius \rf{radius}, and baryonic density \rf{densityq} of the $SU(N)$ False Vacuum Skyrmions with $N=p+q$.

\section{The coupling constant fixing from the experimental data}\label{sec:fixing}
\setcounter{equation}{0}

Now, we show how to fix some of the coupling constants of the model \rf{false0} to compare its results with the experimental data, following the procedure proposed in \cite{luiz:false} for the $SU(2)$ False Vacuum Skyrme model. Additionally, we present how to connect the integration constants $\gamma$ to the experimental data and how to use this  construct topological solutions of the model \rf{false0} for all positive integer values of the topological charge.


The massive term for the $\psi$ field introduced in \rf{potminus1} guarantees that at large distances, our density of nuclear matter ${\cal Q}$ falls exponentially to zero with a rate independent of the mass number. In fact, using \rf{psidef}, \rf{potminus1} and the linearized version of \rf{psieqradial}, we conclude that ${\cal Q}$ falls asymptotically ($r\rightarrow \infty$) as ${\cal Q} \propto e^{-s\,\sqrt{2}\,\frac{\beta_2}{\mu_0}\,r}/r$. Assuming that the experimental baryon density falls exponentially as ${\cal Q}_{{\rm exp.}} \propto e^{-a/r}$, one can use the experimental value of $a$ to fix the ratio $\mu_0 / \beta_2$ through
 \be 
\frac{\mu_0}{\beta_2}=s\,\sqrt{2}\,  a \lab{fixingbeta0mu0}
\ee

Once the parameters $s$ that relate $\psi$ and ${\cal Q}$, and the parameter $\kappa$ of the physical potential $V$ are chosen, we have that $\Lambda\(\gamma,\,s,\,\kappa\)$ is an ordinary functional of the integration constant $\gamma$. Numerically, we can observe that $\Lambda\(\gamma,\,s,\,\kappa\)$ and $I(\gamma, \, s, \,\kappa)$, given respectively in \rf{chargefinal} and \rf{chargefinal}, corresponding to the bounce solutions of \rf{hpsieq}, are monotonically decreasing with $\gamma$. Therefore, since the ratio $\mu_0 / \beta_2$ is fixed through \rf{fixingbeta0mu0}, the rms radius \rf{radius} becomes also an ordinary monotonically decreasing functional of $\gamma$.

It is convenient to use the experimental rms radius of a heavy bound state, denoted by $\sqrt{\langle r \rangle}_{\rm ref.}$, with a mass number $Q_{\rm ref.}$, which lies in a region where the theoretical radius satisfies $\sqrt{\langle r \rangle} \propto Q^{1/3}$, to fix the integration constant corresponding to this reference nucleus, denoted by $\gamma_{\rm ref.}$. Using these parameters and \rf{fixingbeta0mu0}, we can obtain an integral equation for $\gamma_{\rm ref.}$ given by
\be \sqrt{\langle r\rangle}_{{\rm ref.}} = s\,\sqrt{2}\, a\,\Lambda\(\gamma_{{\rm ref.}},\,s,\,\kappa\) \lab{rmsr}
\ee
We must vary the value of the integration constant $\gamma$ of the effective potential until we find the bounce solutions of the reduced equation \rf{qref} that lead to the functional $\Lambda\(\gamma,\,s,\,\kappa\)$ that satisfies \rf{rmsr} to find $\gamma_{{\rm ref.}}$. 

Once $\gamma_{\rm ref.}$ is determined, using \rf{chargefinal}, the values of $\gamma$ corresponding to all other values of the topological charge can be determined through
\be
Q=\frac{Q_{\rm ref.}}{I_{\rm ref.}}\, I\(\gamma,\,s,\,\kappa\) ;\qquad {\rm with } \qquad I_{\rm ref.}=I\(\gamma_{\rm ref.}\,,\,s\,,\,\kappa\) \lab{qref}
\ee
Using the fact that $I\(\gamma,\,s,\,\kappa\)$ is monotonic in $\gamma$, we must vary the value of $\gamma$ until we find the bounce solution of the reduced equation \rf{qref}  that leads to the functional $I\(\gamma,\,s,\,\kappa\)$ that satisfies \rf{qref}. Therefore, to obtain the list of values of $\gamma(Q)$ that lead to positive integer values of $Q$. We compare \rf{qref} with \rf{chargefinal} and, using the fixing of the ratio $\mu_0 / \beta_2$ through \rf{fixingbeta0mu0}, we obtain that the factor $\vartheta$ is determined by
\be
\vartheta \equiv  \(m_0\,e_0\)^3\,\(\frac{\beta_2^2}{\beta_{\kappa}^2}\)^{\frac{s}{\(\kappa-2\)}} = \frac{Q_{\rm ref.}}{I_{\rm ref.}}\,\(\frac{\mu_0}{\beta_2}\)^{-3}   \lab{fixingtheta}
\ee

Note that the topological solutions of \rf{psieqradial} model are constructed by determining the values of the integration constant $\gamma$ that lead to integer values of the topological charge. This is achieved by solving an equation for the $\hpsi$ field, which changes as $\gamma$ varies. Therefore, the integration constant depends on the value of $Q$, similarly to how running coupling constants behave in QCD, but at the classical level.

Curiously, this choice of the reference nucleus and \rf{fixingbeta0mu0} fixes the prediction of the radii of the model without fixing any other coupling constants. Since $\vartheta$ and $\mu_0 / \beta_2$ are fixed, we can obtain $\sigma_2$ and $\sigma_G$ by fitting the binding energy per nucleon given in \rf{binding} to a list of experimental data.  This also fixes the energy term $E_2$ given in \rf{energyf} for all values of $Q$. 

The term $m_0/e_0$ that fixes the contribution of $E_1$ given in \rf{e1}, can be fixed by indentifying the total energy of the model $E=E_1+E_2$ (see \rf{energyf}) with the experimental mass of some state of topological charge $Q=A$, denoted by $m_A^{{\rm exp.}}$, through 
\be \frac{m_0}{e_0}=\frac{m_A^{{\rm exp.}}-E_2\(Q=A\)}{48\,\pi^2\,A} \lab{totalmassf}\ee




The procedure for fixing the coupling constants given above provides five relations for the original six coupling constants of \rf{false0} and fixes all the values of the integration constant for all the nuclei. Choosing the inverse of the dimensionless scale of the field $\psi$, given in \rf{dimensionless}, denoted by $\alpha\equiv \(\beta_\kappa/\beta_2\)^{2/(\kappa-2)}$, as the remaining degree of freedom, we can express the original coupling constants in terms of the fixed quantities $\mu_0 / \beta_2$, $\vartheta$, $\sigma_2$, $\sigma_G$, $m_0 / e_0$, and the free parameter $\alpha$ through
\br
m_0^2&=&\(\frac{m_0}{e_0}\)\,\vartheta^\frac{1}{3}\,\alpha^\frac{s}{3} \;;\qquad e_0^2=\(\frac{m_0}{e_0}\)^{-1}\,\vartheta^\frac{1}{3}\,\alpha^\frac{s}{3}\;;\qquad\frac{\mu_0^2}{\alpha^2}=\frac{\vartheta\,\sigma_2}{4\,\pi^2}\,\(\frac{\mu_0}{\beta_2}\)^2\qquad \label{couplingm}\\
\beta_2^2&=&\frac{\vartheta\,\sigma_2}{4\,\pi^2}\,\alpha^2\;;\qquad\qquad \:\,\beta_\kappa^2=\frac{\vartheta\,\sigma_2}{4\,\pi^2}\,\alpha^\kappa \;;\qquad\qquad\quad \quad\beta_G^2=\frac{\vartheta\,\sigma_2\,\sigma_G}{4\,\pi^2}\,\alpha^s\nonumber
\er

The approach above to construct a classical field theory of topological solitons in agreement with the experimental data is quite surprising, since it depends only on the determination of the topological charge density ${\cal Q}$. This is much simpler than having to solve \rf{psig} to obtain the profile function before comparing the results with the experimental data. In fact, the explicit form of the profile function is not needed to compute the rms radius, the total energy density, the topological charge density, and so on.

\section{Aplication to nuclear physics for the case $G=SU(2)$}
\label{sec:su(2)app}
\setcounter{equation}{0}

The application of the $G = SU(2)$ False Vacuum Skyrme model to nuclear physics provides a good description of several bulk properties of nuclei, such as the binding energy and rms radii, provided the nuclei are sufficiently heavy, as demonstrated in \cite{luiz:false}. However, our goal in this section is to investigate the accuracy of the theoretical baryon density, particularly its value at the center of the nucleus, and its dependence on the exponent $\kappa$ of the highest-order term in the potential \rf{pot0}. We employ the same numerical approach used in \cite{luiz:false} to find the bounce solutions of \rf{hpsieq}, based on the shooting method and the fourth-order explicit Runge-Kutta method with a step size of $\Delta \zeta = 10^{-4}$.

In this paper we consider the same list of $N_c=265$ nuclei presented in \cite{luiz:false}. Its defined by all the stable nuclei according to \cite{nubase2016} including isobars, and above the mass of $^{208}$Pb, which is the heavier stable nuclei, we also consider all the nuclei with a half-life greater than $10^3$ years. Inside these criteria $^{240}$Pu is the heavier nuclei and there is a gap in the mass number $A$ from $A=210$ to $A=225$.  For this list of nuclei the experimental values of charge radii as given by \cite{angeli}, and the values of binding energies per nucleon as given in \cite{ame2016}. 

The theoretical results are compared with the experimental data of the list of nuclei defined above using RMSD of the rms radii $R\equiv \sqrt{\langle r^2\rangle}$ and the binding energy per nucleon $E_B$, defined respectively by
\be \Delta\,R=\sum_{{\rm list}}\sqrt{\frac{\(R^{{\rm num.}}-R^{{\rm exp.}}\)^2}{N_c}};\qquad\qquad \Delta\,E_B=\sum_{{\rm list}}\sqrt{\frac{\(E_B^{{\rm num.}}-E_B^{{\rm exp.}}\)^2}{N_c}} \lab{rmsd}
\ee
where $R^{{\rm exp.}}$ and $E_B^{{\rm exp.}}$ are the rms radii and binding energy per nuclei, according to given respectively to \cite{angeli} and \cite{ame2016}. In addition, $R^{{\rm num.}}$  and $E^{{\rm num.}}$ are the numerical values obtained respectively from \rf{radius} and \rf{binding}. For excluding the nine nuclei lightest than $^{12}$C in our list, we also define $\Delta\,R^{A \geq 12}$ and $\Delta\,E_B^{A \geq 12}$ by restricting the sum of \rf{rmsd} to $A \geq 12$ and taking $N_c^{A \geq 12}=256$. 

We follow the procedure described in Section \ref{sec:fixing} to fix the coupling constants, also followed in \cite{luiz:false}. For sufficiently heavy nuclei, the experimental baryon density is well described by the two-parameter Fermi model, taking the form ${\cal Q}\(r\)={\cal Q}\(0\)/\(1+e^{\(r-c\)/a}\)$, where $c$ and $a$ are constants, with $a = 0.524\,{fm}$. Therefore, the ratio $\mu_0/\beta_2$ is fixed according to \rf{fixingbeta0mu0}. 

As shown in \cite{luiz:false}, the rms radius and the binding energy per nucleon does not strongly depend on the choice of the reference nucleus, as long as it is heavy enough. Therefore, we can choose the $^{56}$Fe as the reference nucleus, which implies that $Q_{{\rm ref.}}=56$ and $\sqrt{\langle r\rangle}_{{\rm ref.}}=3.7377\,fm $ (see \rf{rmsr} and \rf{qref}). The construction of solutions to \rf{hpsieq} satisfying \rf{qref} for $s = 2.5,\, 3,\, 5$ and $\kappa = 4,\,\ldots,\, 20$ reveals that for $s = 3$, the rms radius reproduces the experimental data with good accuracy as long as $\kappa \geq 6$ and $Q \geq 12$, where $\Delta R$ and $\Delta R^{A \geq 12}$, defined in \rf{rmsd}, are approximately $0.08\,{fm}$ and $0.04\,{fm}$, respectively. This provides a physical motivation to set the parameter to $s = 3$ in our model, where following the approach proposed in \cite{luiz:false}, $\Delta E$ and $\Delta E^{A \geq 12}$ for $4 \leq \kappa \leq 20$ are about $0.2\,{MeV}$ and $0.05\,{MeV}$, respectively.

In the $SU(2)$ case, the isospin symmetry ensures that the total mass of a nucleus increases approximately linearly with the number of nucleons, since the binding energy constitutes only about $1\%$ of the nuclear mass. Consequently, one does not expect a substantial difference in fixing the ratio $m_0/e_0$ using a heavy nucleus versus a light one, such as the proton ($^1$H). This rationale motivates the natural choice of calibrating $m_0/e_0$ by identifying the total energy of the $Q=1$ False Vacuum Skyrmion with the proton mass $m_p$, i.e., by setting $A=1$ and $m_A^{\rm exp.} = m_p$ in \rf{totalmassf}, where where $m_p=938.27208816\, MeV$, according to \cite{ParticleDataGroup:2024cfk}.




In \cite{luiz:false}, the parameters $\sigma_2$ and $\sigma_G$ that minimize $\Delta E$ given in \rf{rmsd} were determined for $s = 3$ and $\kappa = 4,\,\ldots,\,20$, following the procedure described in Section \ref{sec:fixing} that fixes the coupling constants. In this work, we first follow that approach to extend those results up to $\kappa = 25$. Then, we adopt an alternative approach in which we find the values of the parameters $\sigma_2$ and $\sigma_G$ that minimize $\Delta E^{A \geq 12}$, since for light nuclei ($A < 12$) the accuracy of our model is very low.

The fitting of the two parameters $\sigma_2$ and $\sigma_G$ affects neither the RMS radius nor the topological charge density by construction, as explained in Section \ref{sec:fixing}. Therefore, for $\kappa = 4, \,...,\, 20$, we obtain exactly the same results as those in \cite{luiz:false} for the RMS radius. However, our new approach reduces the values of $\Delta E^{A \geq 12}$ by about $8.29-8.91\,\%$, while $\Delta E$ increases by only about $0.40-0.43\,\%$ for $\kappa = 4, \ldots, 25$ and $s = 3$ (see Table \ref{table}).

\begin{longtable}{@{\extracolsep{\fill}}c c c c c c c@{}}
\caption{The RMSD of the RMS radius $\Delta R\;(fm)$ and $\Delta R^{A \geq 12}\;(fm)$, and binding energy per nucleon $\Delta E_B\;(MeV)$ and $\Delta E_B^{A \geq 12}\;(MeV)$, as defined in \rf{rmsd}, obtained in \cite{luiz:false} by minimizing $\Delta E_B$ for $\kappa=4,\dots,20$ with $s=3$. We extend the results presented in the second to fifth columns up to $\kappa=25$ by employing the same approach. The sixth and seventh columns show the values of $\Delta E_B$ and $\Delta E_B^{A \geq 12}$, respectively, obtained by minimizing $\Delta E_B^{A \geq 12}$ for the same values of $s$ and $\kappa$.} 
\label{table} \\
\hline
$\kappa$ & $\Delta R$ \cite{luiz:false}  & $\Delta R^{A\geq 12}$ \cite{luiz:false}  & $\Delta E_B$ \cite{luiz:false} & $\Delta E_B^{A\geq 12}$ \cite{luiz:false} & $\Delta E_B$ & $\Delta E_B^{A\geq 12}$ \\
\hline
\endfirsthead

\hline
$\kappa$ & $\Delta R$ \cite{luiz:false}  & $\Delta R^{A\geq 12}$ \cite{luiz:false}  & $\Delta E_B$ \cite{luiz:false} & $\Delta E_B^{A\geq 12}$ \cite{luiz:false} & $\Delta E_B$ & $\Delta E_B^{A\geq 12}$ \\
\hline
\endhead

$4$ & $0.128737$ & $0.101077$ & $0.215518$ & $0.047759$ & $0.216389$ & $0.043751$ \\
$5$ & $0.093324$ & $0.056445$ & $0.216087$ & $0.047923$ & $0.216999$ & $0.043722$ \\
$6$ & $0.084536$ & $0.043729$ & $0.216271$ & $0.048066$ & $0.217200$ & $0.043796$ \\
$7$ & $0.082406$ & $0.040774$ & $0.216295$ & $0.048156$ & $0.217230$ & $0.043866$ \\
$8$ & $0.082042$ & $0.040475$ & $0.216247$ & $0.048213$ & $0.217183$ & $0.043923$ \\
$9$ & $0.082159$ & $0.040788$ & $0.216166$ & $0.048250$ & $0.217101$ & $0.043972$ \\
$10$ & $0.082391$ & $0.041176$ & $0.216071$ & $0.048275$ & $0.217003$ & $0.044014$ \\
$11$ & $0.082628$ & $0.041508$ & $0.215972$ & $0.048292$ & $0.216900$ & $0.044051$ \\
$12$ & $0.082841$ & $0.041767$ & $0.215873$ & $0.048306$ & $0.216797$ & $0.044086$ \\
$13$ & $0.083022$ & $0.041962$ & $0.215777$ & $0.048317$ & $0.216698$ & $0.044117$ \\
$14$ & $0.083174$ & $0.042105$ & $0.215686$ & $0.048326$ & $0.216604$ & $0.044147$ \\
$15$ & $0.083301$ & $0.042211$ & $0.215600$ & $0.048334$ & $0.216514$ & $0.044174$ \\
$16$ & $0.083407$ & $0.042287$ & $0.215520$ & $0.048342$ & $0.216430$ & $0.044200$ \\
$17$ & $0.083496$ & $0.042343$ & $0.215445$ & $0.048349$ & $0.216352$ & $0.044225$ \\
$18$ & $0.083571$ & $0.042382$ & $0.215375$ & $0.048355$ & $0.216279$ & $0.044248$ \\
$19$ & $0.083635$ & $0.042410$ & $0.215310$ & $0.048362$ & $0.216211$ & $0.044270$ \\
$20$ & $0.083689$ & $0.042429$ & $0.215249$ & $0.048368$ & $0.216147$ & $0.044291$ \\
$21$ & $0.083735$ & $0.042441$ & $0.215192$ & $0.048374$ & $0.216088$ & $0.044311$ \\
$22$ & $0.083775$ & $0.042448$ & $0.215139$ & $0.048379$ & $0.216033$ & $0.044330$ \\
$23$ & $0.083809$ & $0.042452$ & $0.215090$ & $0.048385$ & $0.215981$ & $0.044348$ \\
$24$ & $0.083838$ & $0.042453$ & $0.215044$ & $0.048390$ & $0.215933$ & $0.044365$ \\
$25$ & $0.083864$ & $0.042451$ & $0.215001$ & $0.048395$ & $0.215887$ & $0.044382$ \\
\hline
\end{longtable}

In order to exemplify the behaviour and the construction of the False Vacuum Skyrmions, let us consider the case with $s = 3$ and $\kappa = 25$. Once the parameters $s$ and $\kappa$ are fixed, and the parameter $a$ is given by \rf{fixingbeta0mu0}, the integration constant $\gamma_{^{56}Fe}$ can be obtained by solving \rf{rmsr} numerically. Then, the ratio $\frac{Q_{\rm ref.}}{I_{\rm ref.}}$ is fixed, and the other values of $\gamma$ corresponding to integer values of $Q$ are determined by solving \rf{qref}. In both cases, it is necessary to find the bounce solutions of \rf{hpsieq} that satisfy these equations

\begin{figure}[htp!]
\begin{center}
		\includegraphics[scale=0.47]{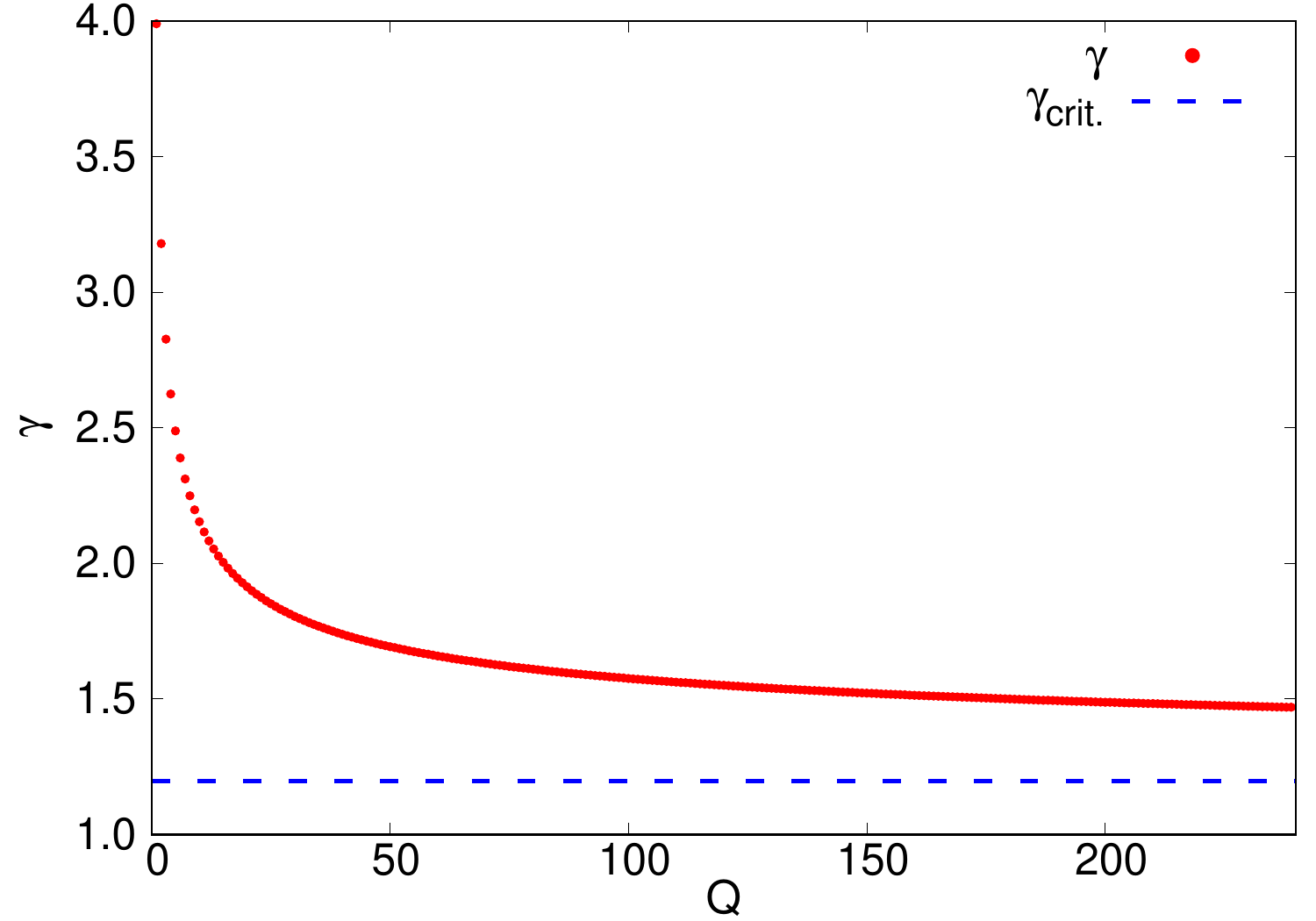}
		\caption{The integration constant $\gamma$ as a function of $Q$  and its critical value $\gamma_{{\rm crit.}}$ given by \rf{critical} for $s=3$ and  $\kappa=25$.}
		\label{fig:gamma}
\end{center} 
\end{figure}

For $s = 3$ and $\kappa = 25$, the integration constant decreases from $\gamma = 3.99052$ at $Q = 1$ to $\gamma = 1.46904$ at $Q = 240$, varying more rapidly for light nuclei and more slowly for heavier nuclei, as shown in Figure \ref{fig:gamma}. Using the fact that the length scale $\frac{\mu_0}{\beta_2}$ is fixed through \rf{fixingbeta0mu0}, the parameter $\vartheta$ is determined by \rf{fixingtheta} and corresponds to $\vartheta = 8.31937$. Therefore, the nuclear density \rf{nucleardensity} is fully determined, even without fitting the additional parameters $\sigma_2$ and $\sigma_G$ related to the binding energy per nucleon \rf{binding}.

Heavier nuclei can be constructed for lower values of $\gamma$ than those shown in Figure \ref{fig:gamma}. Infinitely heavy nuclei correspond to the limit $\gamma \rightarrow \gamma_{\rm crit.} = 1.19584$ (see \rf{critical}) for $s=3$ and $\kappa=25$, which implies $Q \rightarrow \infty$, as explained in Section \ref{sec:euler}. Therefore, all nuclei from $Q = 240$ up to $Q \rightarrow \infty$ can be constructed within a narrow range of $\gamma$, where $\gamma_{\rm crit.}$ is only about $18.60\%$ lower than $\gamma_{Q=240}$.

Table \ref{table} presents the values of $\Delta R$ and $\Delta R^{A \geq 12}$, as defined in \rf{rmsd}, which exhibit only minor variation for $\kappa \geq 6$. For those values of $\kappa$, $\Delta R$ ranges from $0.082042\, fm $ to $0.084536\, \mathrm{fm}$, while $\Delta R^{A \geq 12}$ ranges from $0.040475\, fm$ and $0.043729\, fm$. These values are small when compared with the typical experimental sizes of the nuclei, which range from $0.8783\,fm$ to $5.8701\,fm$ in our list of nuclei. 

\begin{figure}[htp!]
\begin{center}
		\includegraphics[scale=0.47]{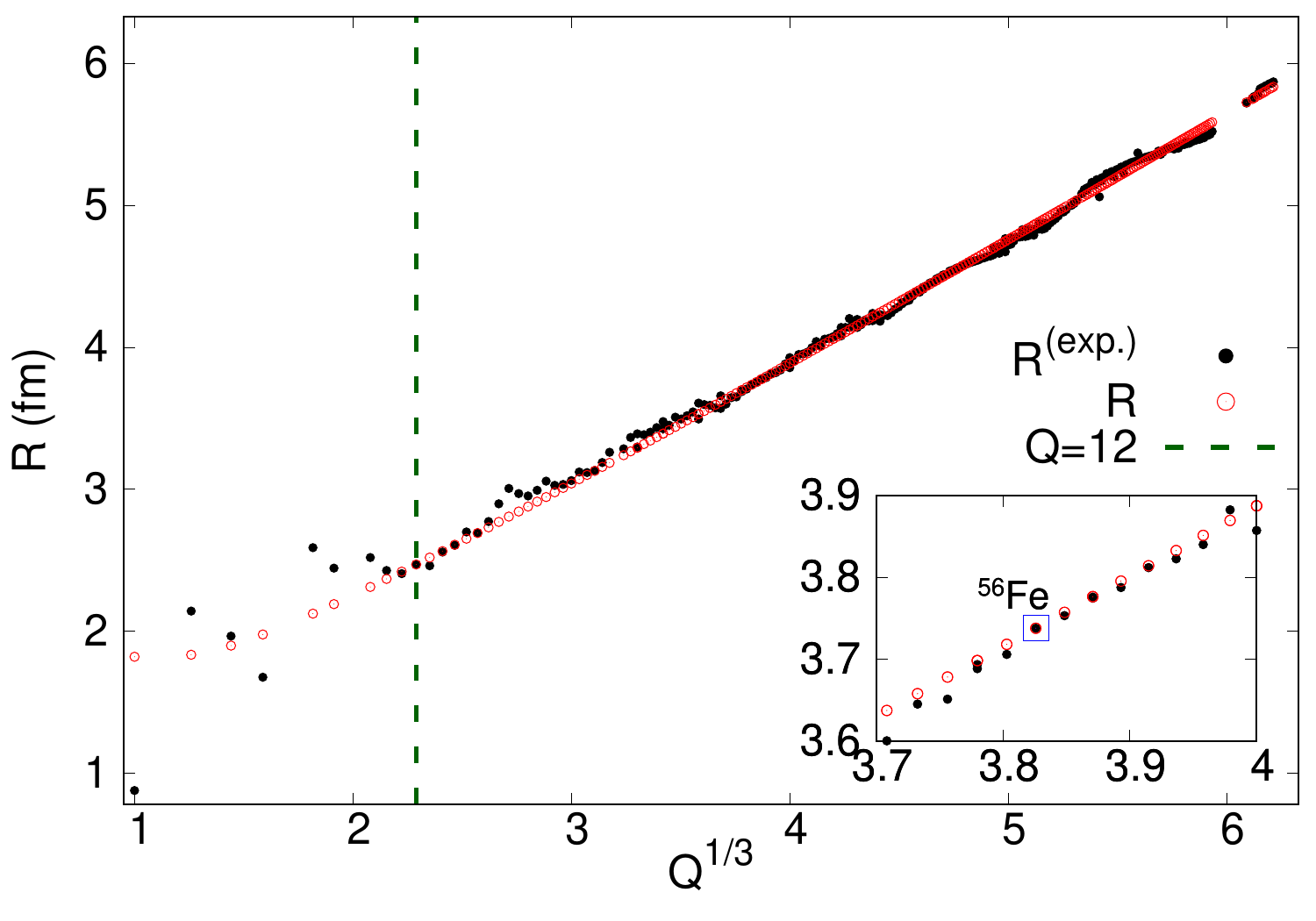}
		\caption{The RMS radii $R\equiv \sqrt{\langle r^2\rangle}$ as a function of $Q^{1/3}$ for $s=3$ and $\kappa=25$.}
		\label{fig:radii}
\end{center} 
\end{figure}

Figure \ref{fig:radii} shows the experimental RMS radii \rf{radius} of all $265$ nuclei in our list, ranging from $A = 1$ to $240$, as defined in Section \ref{sec:fixing}, along with the corresponding theoretical results obtained from \rf{radius} for $s = 3$ and $\kappa = 25$. The mass number $A$ is identified with the topological charge, i.e., $A = Q$, and the experimental data include isobars. Although the model \rf{false0} has solutions for each integer value of the topological charge within this mass range, we present numerical results only for those values of $Q$ for which there is at least one nucleus in the experimental dataset with the same mass number. 

The gaps in the data shown in Figure \ref{fig:radii}  correspond to those values of $Q$ for which there are no experimental nuclei that satisfy our selection criteria based on stability and half-life.  The theoretical radius reproduces the experimental results with good accuracy, provided that the nucleus is sufficiently heavy, which we define by the mass number condition $A \geq 12$, where $\sqrt{\langle r^2 \rangle} \sim Q^{1/3}$.

\begin{figure}[htp!]
\begin{center}
		\includegraphics[scale=0.47]{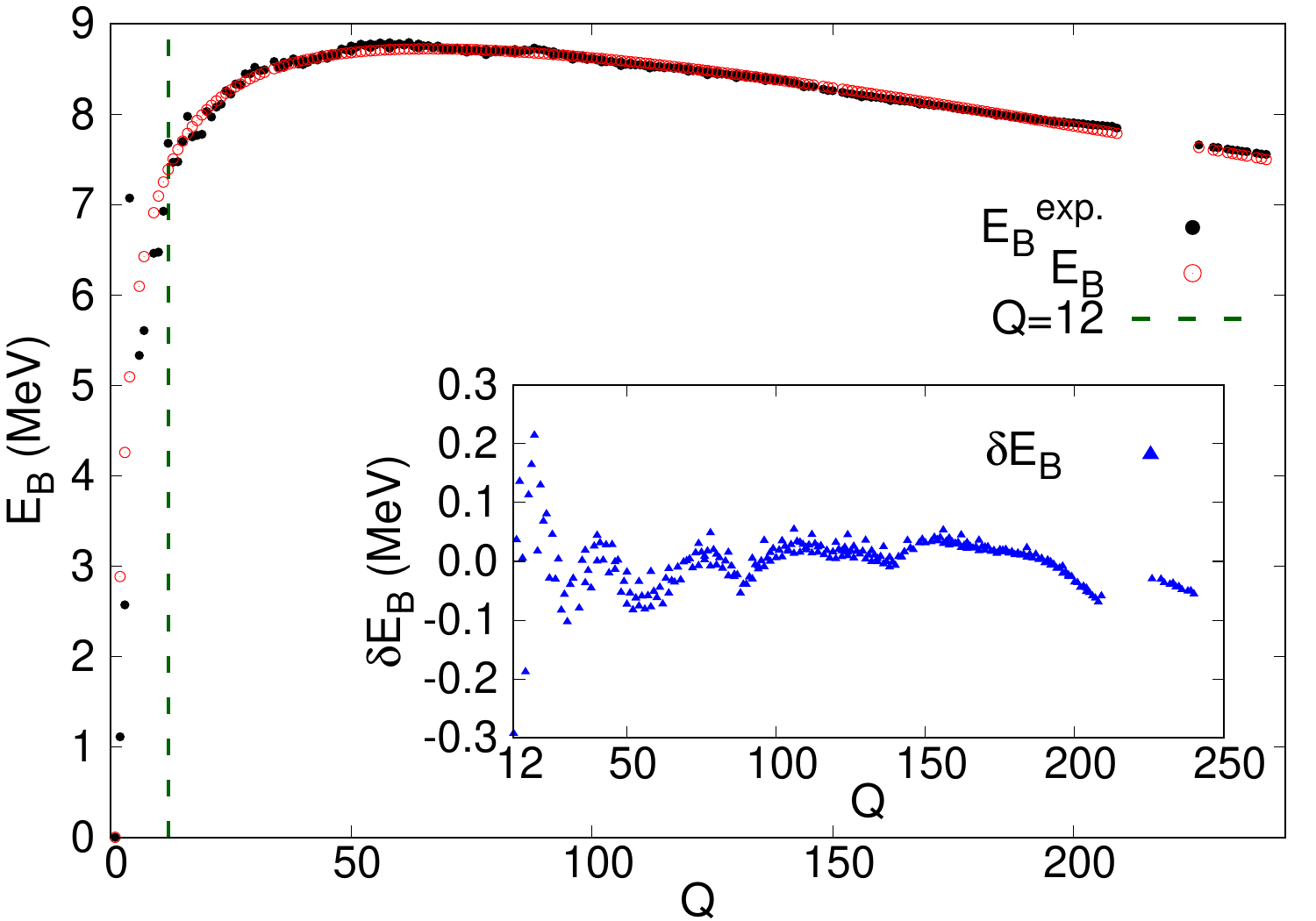}
		\caption{The binding energy per nucleon $E_B$ and $\delta E_B$ as a function of $Q$ for $s=3$ and $\kappa=25$, corresponding to the minima of $\Delta E_B^{A\geq 12}$ given by $\sigma_2=2.34910$ and $\sigma_G=9.57606 \times 10^{-3}$, where $\Delta E_B^{A\geq 12} =0.044382\, MeV$. }
		\label{fig:binding}
\end{center} 
\end{figure}

Figure \ref{fig:binding} shows that, at least in the region $A \geq 12$, our model provides a good description of the binding energy per nucleon $E_B$ for $s=3$ and $\kappa=25$. In fact, the value of $\Delta E^{A \geq 12}$ corresponds to $0.65\%$ of the lowest experimental value of the binding energy per nucleon $E_B^{\mathrm{exp.}}\left(^{13}\mathrm{C}\right) = 7.469849\,\mathrm{MeV}$ in this restrictive list of $256$ nuclei, while it corresponds to $0.55\%$ of the highest experimental value of the binding energy per nucleon $E_B^{\mathrm{exp.}}\left(^{56}\mathrm{Fe}\right) = 8.790354\,\mathrm{MeV}$. In addition, Figure \ref{fig:binding} also shows the values of $\delta E_B \equiv E_B - E_B^{\text{exp.}}$ for $A \geq 12$, where the largest absolute value of $\delta E_B$ is $0.292646\,\mathrm{MeV}$ at $Q=12$. Restricting the analysis to even heavy nuclei with $A \geq 31$, the maximum of $\mid \delta E_B\mid$ becomes $0.082763\,\mathrm{MeV}$ at $Q = 52$. Since this value decreases by approximately a factor of $3.54$, the accuracy in describing the binding energy per nucleon is substantially improved in this region when compared with the region $12 \leq A \leq 30$.

A fraction of the RMSD in our binding energy per nucleon arises from the fact that our list of nuclei includes isobars, while our model does not distinguish between protons and neutrons. In order to estimate this effect over the total RMSD, consider an ideal function $E_B^{{\rm (ideal)}}(Q)$ that depends only on $Q$ and minimizes the RMSD $\Delta E$. The value of $E_B^{{\rm (ideal)}}(Q)$ must correspond to the median value of the experimental binding energy per nucleon for each fixed value of $Q$. For our list of nuclei, the RMSD corresponds to $\Delta E_B^{{\rm (ideal)}} =8.27966\,keV$ and $\Delta\,E_B^{{\rm (ideal)}\,A\geq 12}=8.42395\,keV$, which are respectively about $3.84\,\%$ and $19.98\,\%$ of the RMSD of our model for $s=3$ and $\kappa=25$.

\begin{figure}[htp!]
\begin{center}
		\includegraphics[scale=0.47]{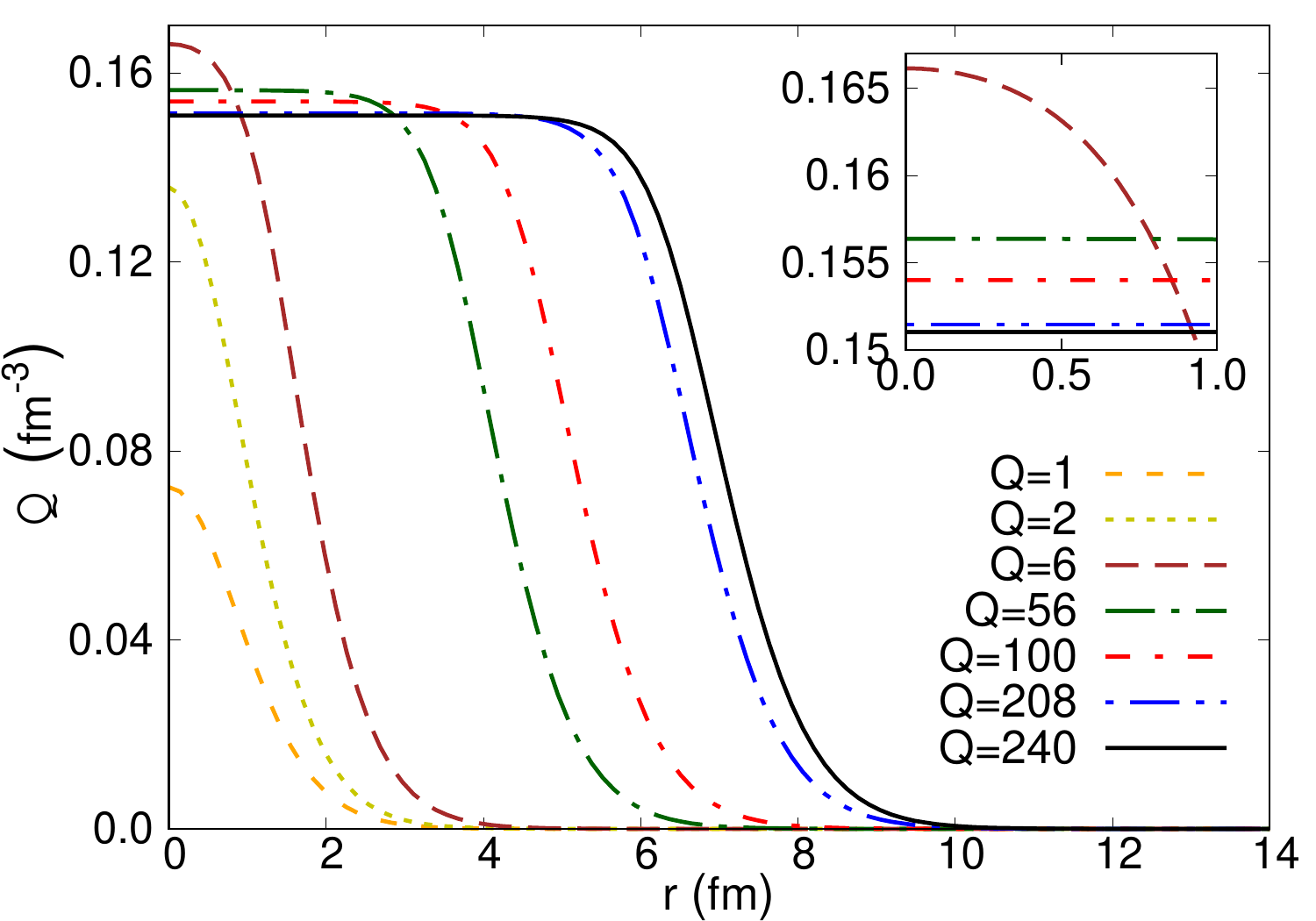}
			\caption{The topological charge density ${\cal Q} = \frac{\vartheta}{4\,\pi^2}\hpsi^s\(\zeta=r/\(s\,\sqrt{2}\,0.524\,fm\)\)$ for $s=3$ and $\kappa=25$ using we method of our paper. Note that the value of ${\cal Q}\(0\)$ increases up to $Q = 6$ and then decreases slowly to $0.151023\; fm^{-3}$ for $Q=240$.}
		\label{fig:density}
\end{center} 
\end{figure}

The topological charge density for $s=3$ and $\kappa=25$ is shown in Figure \ref{fig:density} for some values of $Q$, and it approximates a Woods-Saxon distribution, as expected. The nuclear density at the centre of the nuclei ${\cal Q}(0)$, obtained through \rf{criticaldensity} at $r=0$, rises for low values of $Q$, reaches a maximum at $Q=6$ where ${\cal Q}(0)= 0.166141 \,\mathrm{fm}^{-3}$, and then decreases slowly for larger values of $Q$, reaching ${\cal Q}(0)= 0.151023 \,\mathrm{fm}^{-3}$ at $Q=240$. This behaviour is unexpected from an experimental standpoint, as ${\cal Q}(0)$ is expected to remain approximately constant with increasing $Q$. However, this behaviour is independent of the values of $\kappa$ for $\kappa = 4, \, ...,\,  25$, although the specific values of ${\cal Q}(0)$ and the value of $Q$ at which ${\cal Q}(0)$ reaches its maximum may vary, as shown in Table \ref{table2}.

The theoretical pattern of ${\cal Q}(0)$ observed in Figure \ref{fig:density} as a function of $Q$ is expected, since ${\cal Q}(0)$ is directly proportional to $\hpsi^s(\zeta=0)$. For a fixed values of $s$ and $\kappa$, $\hpsi^s(\zeta=0)$ can be written as a ordinary function of the integration constant or alternativelt the $I$ functional, where $I$ is itself directly proportional to the topological charge (see equation \rf{chargefinal}). As shown in \cite{luiz:false}, $\hpsi(I)$ at $\zeta = 0$ increases monotonically for small values of $I$, until it reaches its maximum at some value $I = I_m$, which depends on $s$ and $\kappa$, and for $I > I_m$ exhibits a very slow, monotonic decay.

The value $Q_m$ corresponding to the topological charge at which ${\cal Q}(0)$ reaches its maximum, for a fixed value of $\kappa$, and its associated nuclear density at the origin ${\cal Q}_m(0)$ are given in Table \ref{table2}. Additionally, this table includes the parameter $\vartheta$ obtained from \rf{fixingtheta}, the critical nuclear density at the origin ${\cal Q}_{{\rm crit.}}(0)$ given in \rf{criticaldensity}, as well as the values of ${\cal Q}_{Q=208}(0)$, ${\cal Q}_{Q=240}(0)$, and the difference $\Delta {\cal Q}(0) \equiv {\cal Q}_m(0) - {\cal Q}_{Q=240}(0)$. 

\begin{figure}[htp!]
\begin{center}
		\includegraphics[scale=0.47]{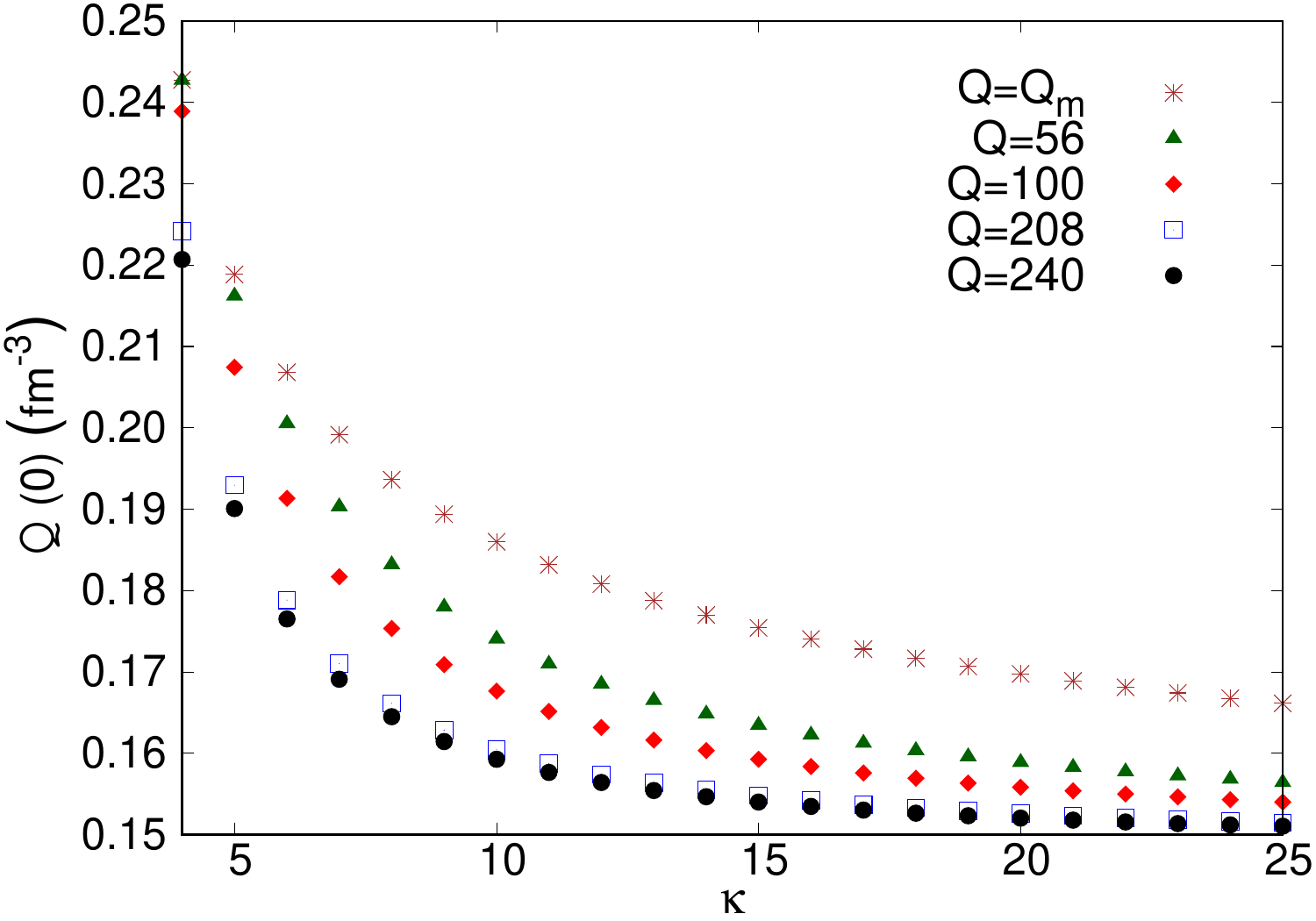}
			\caption{The topological charge density at the centre of the nuclei ${\cal Q}(0)$ for $\kappa=4,\,...,\, 25$ and $s=3$.}
		\label{fig:asymp}
\end{center} 
\end{figure}

The value of $Q_m$ decreases monotonically with $\kappa$ and becomes small for large values of $\kappa$. Indeed, for all $\kappa \geq 11$, we find that $Q_m \leq 12$. Consequently, throughout the entire domain corresponding to our restricted set of nuclei with mass number $A \geq 12$, the quantity ${\cal Q}(0)$ exhibits a monotonically decreasing behaviour as a function of $Q$. On the other hand, Figure \ref{fig:asymp} shows that, for each fixed value of the topological charge $Q=56,\, 100,\,208,\,240$, the nuclear density at the origin ${\cal Q}(0)$ is monotonically decreasing with $\kappa$ for $\kappa=4,\, ... ,\, 25$.

For lower values of $\kappa$, the values of ${\cal Q}_{Q=208}(0)$ differ significantly from the interior nuclear density of the $^{208}$Pb nucleus ${\cal Q}_{^{208}{\rm Pb}}^{{\rm exp.}}=0.1480 \pm 0.0038\, fm^{-3}$, according to \cite{prex}. However, this difference drops monotonically with $\kappa$. Thus, the theoretical predictions for ${\cal Q}(0)$ improve significantly with increasing $\kappa$, as the error decreases from $10.03\%$ at $\kappa=9$ to $2.34\%$ at $\kappa=25$. In addition, this error is less than or equal to $3.10\%$ as long as $\kappa \geq 20$. On the other hand, increasing the value of $\kappa$ also maintains the good accuracy of the RMS radius and binding energy per nucleon, since the values of $\Delta R^{A \geq 12}$ and $\Delta E^{A \geq 12}$ remain small for $\kappa \geq 6$ (see Table \ref{table}).

\begin{figure}[htp!]
\begin{center}
		\includegraphics[scale=0.47]{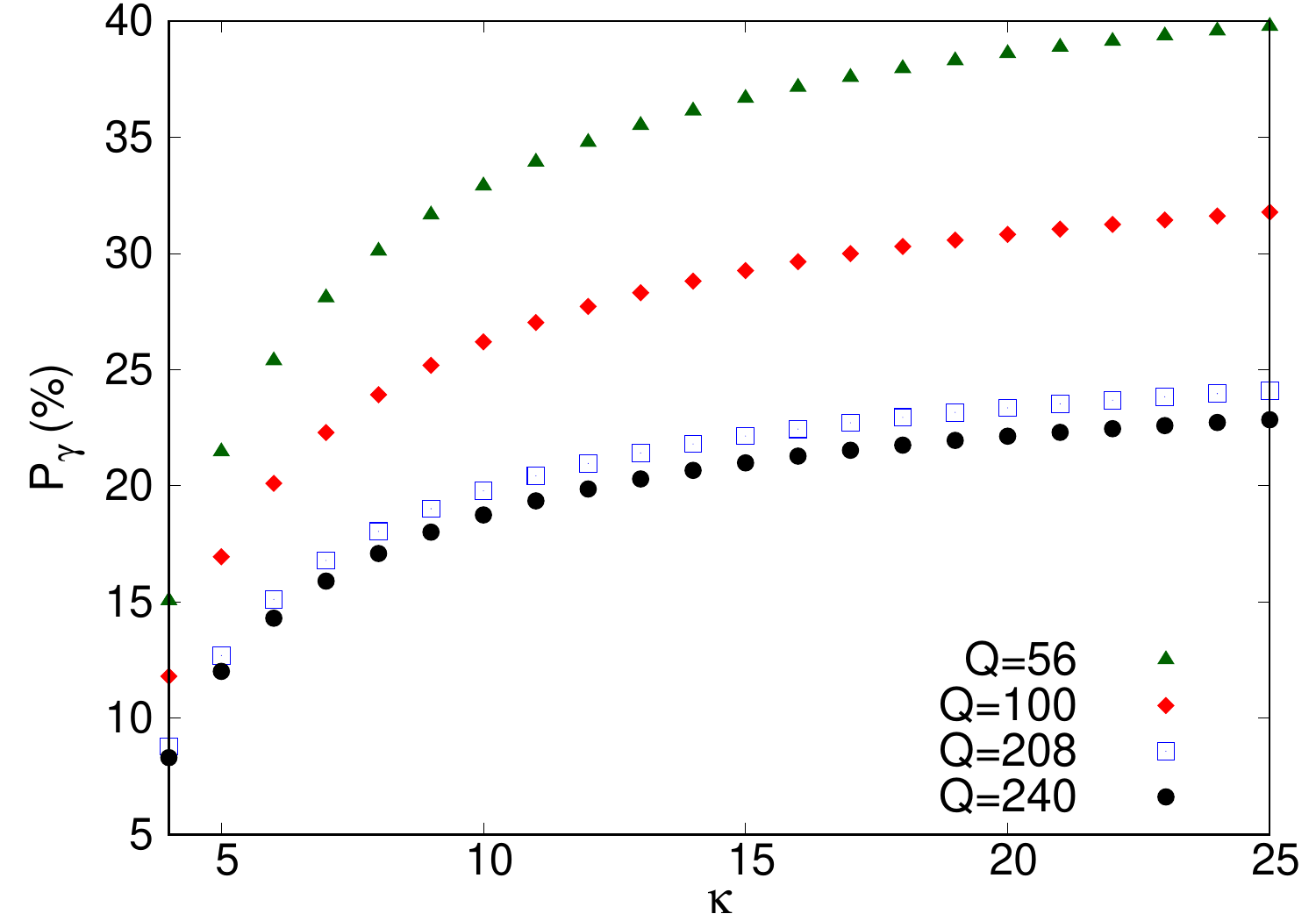}
			\caption{The ratio $P_\gamma = ( \gamma_Q - \gamma_{{\rm critic.}}) / \gamma_{{\rm critic.}} $ in percentage for $\kappa=4,\,...,\, 25$, $Q=56,\, 100,\,208,\,240$ and $s=3$.}
		\label{fig:perc}
\end{center} 
\end{figure}

Table \ref{table2} shows that  for $s=3$ the critical nuclear density at the origin ${\cal Q}_{{\rm crit.}}(0)=\frac{\vartheta}{4\,\pi^2}\,\hpsi^s_{{\rm crit.}}$, as defined in \rf{criticaldensity}, increases monotonically with $\kappa$ for $\kappa \geq 8$. Therefore, in this region, the monotonic decrease of $\vartheta$ with $\kappa$, shown in Table \ref{table2}, is more than compensated by the monotonic increase of $\hpsi_{{\rm crit.}}^3$, as given by \rf{critical}. Within this range of $\kappa$ values, the nuclear densities at the origin for heavy nuclei with $Q = 56,\, 100,\, 208,\, 240$ decrease monotonically with $\kappa$, thereby approaching the critical density as $\kappa$ increases. However, this does not mean that for those values of $Q$ the integration constant is getting closer to its critical value given by \rf{critical}. In fact, we observe the opposite. The relative difference of $\gamma_Q$ with respect to $\gamma_{{\rm critic.}}$, defined as $P_\gamma \equiv (\gamma_Q - \gamma_{{\rm critic.}})/\gamma_{{\rm critic.}}$, grows monotonically with $k$ for $k = 4,\, ...,\, 25$ for each value of $Q = 56,\, 100,\, 208,\, 240$, as shown in Figure \ref{fig:perc}.

\begin{figure}[htp!]
\begin{center}
		\includegraphics[scale=0.47]{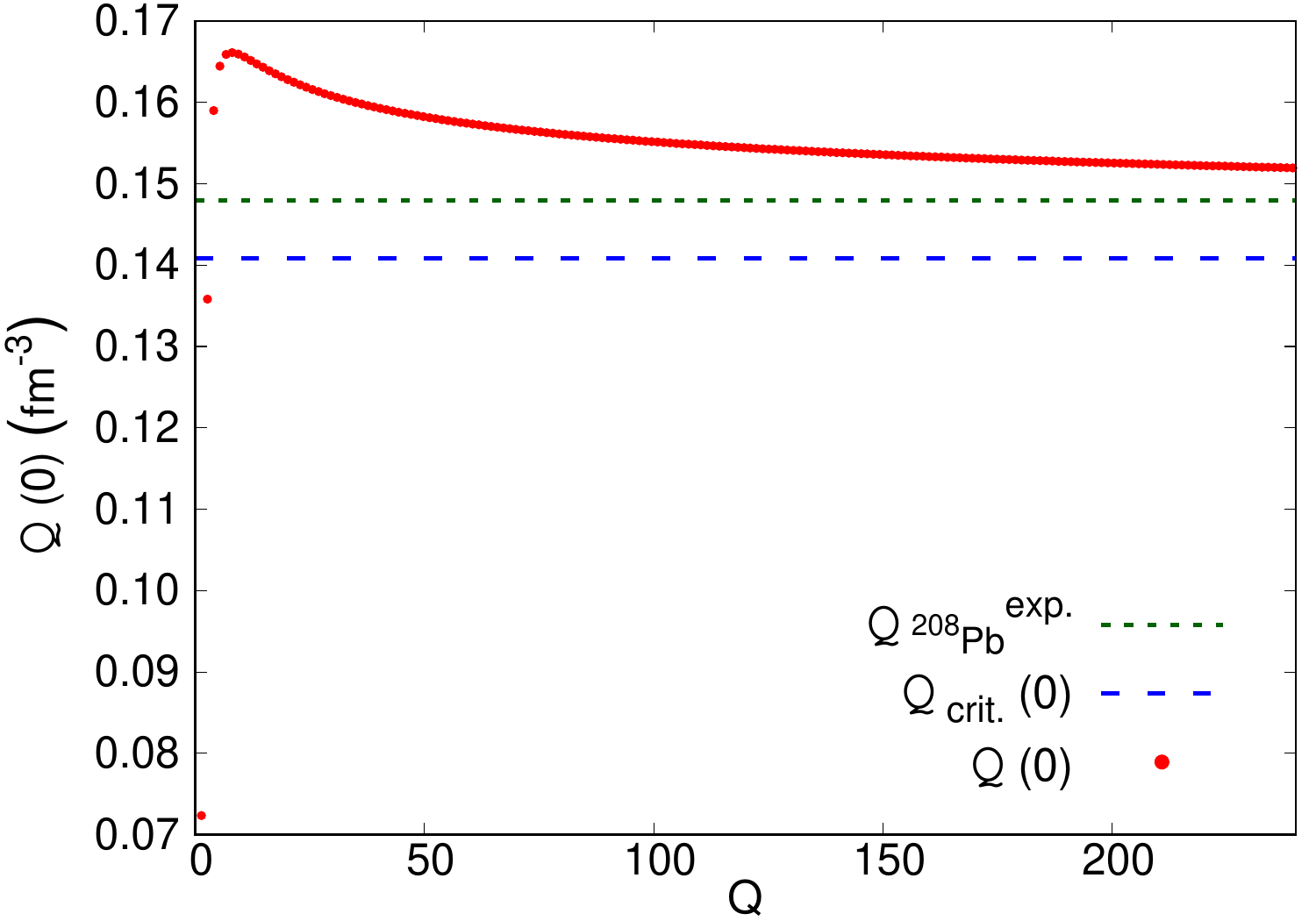}
		\caption{The nuclear density at the origin $\mathcal{Q}(0)$ (blue dots) shown as a function of $Q$, along with the critical value $\mathcal{Q}_{\rm crit.}(0)$ (dashed green line) for $\kappa = 25$ and $s = 3$. In addition, the experimental value of the interior nuclear density of $^{208}$Pb given by $\mathcal{Q}_{^{208}{\rm Pb}}^{\rm exp.}= 0.1480\,fm^{-3}$ (dashed red line) is also presented.}
		\label{fig:density2}
\end{center} 
\end{figure}

Figure \ref{fig:asymp} and Table \ref{table2} also show that the difference $\Delta Q(0)$ grows with $\kappa$ for $\kappa=4,\,5,\,6$, and is monotonically decreasing with $\kappa$ for $\kappa \geq 6$. Since a large variation of ${\cal Q}(0)$ is not expected experimentally, this also shows that our model better reproduces the experimental behaviour of the nuclear density for large values of $\kappa$. In fact, the maximum $(\kappa=6)$ and minimum $(\kappa=25)$ values of variation $\Delta {\cal Q}(0)$ correspond to approximately $20.50\,\%$ and $10.21\,\%$ of ${\cal Q}_{^{208}{\rm Pb}}^{{\rm exp.}}$, respectively. The fact that $\Delta \mathcal{Q}(0)$ is small for $s = 3$ and $\kappa = 25$, and that its values approach the experimental interior nuclear density ${\cal Q}_{^{208}{\rm Pb}}^{{\rm exp.}}$, can also be observed in Figure \ref{fig:density2}, which shows the theoretical values of $\mathcal{Q}(0)$ for $Q = 1$ up to $Q = 240$ for these parameters.

\begin{longtable}{@{\extracolsep{\fill}}c c c c c c c c@{}}
\caption{The parameter $\vartheta$, obtained from \rf{fixingtheta}, and the value of the topological charge $Q_{m}$ corresponding to the topological solution with the highest nuclear density at the origin, are presented for $\kappa = 4, \ldots, 25$ and $s = 3$, following the approach described in Section \ref{sec:fixing}, using $^{56}$Fe as the reference nucleus. In addition, its also presented ${\cal Q}_m(0)$, ${\cal Q}_{Q=240}(0)$, $\Delta {\cal Q}(0)$, ${\cal Q}_{{\rm crit.}}(0)$, and ${\cal Q}_{Q=208}(0)$ in units of ${fm}^{-3}$.} \label{table2} \\

\hline
$\kappa$ & $\vartheta$ & $Q_{m}$ & ${\cal Q}_m(0)$ & ${\cal Q}_{Q=240}(0)$ & $\Delta {\cal Q}(0)$ & ${\cal Q}_{{\rm crit.}}(0)$ & ${\cal Q}_{Q=208}(0)$ \\
\hline
\endfirsthead

\hline
$\kappa$ & $\vartheta$ & $Q_{m}$ & ${\cal Q}_m(0)$ & ${\cal Q}_{Q=240}(0)$ & $\Delta {\cal Q}(0)$ & ${\cal Q}_{{\rm crit.}}(0)$ & ${\cal Q}_{Q=208}(0)$ \\
\hline
\endhead

\hline
\endfoot

\hline
\endlastfoot

$4$ & $5.11903$ & $60$ & $0.242788$ & $0.220699$ & $0.022089$ & $0.129667$ & $0.224213$ \\ 
$5$ & $10.01702$ & $35$ & $0.218871$ & $0.190078$ & $0.028793$ & $0.126867$ & $0.192960$ \\ 
$6$ & $11.47362$ & $25$ & $0.206841$ & $0.176508$ & $0.030333$ & $0.127497$ & $0.178834$ \\ 
$7$ & $11.68246$ & $20$ & $0.199160$ & $0.169086$ & $0.030074$ & $0.128807$ & $0.171009$ \\ 
$8$ & $11.49218$ & $17$ & $0.193635$ & $0.164498$ & $0.029136$ & $0.130184$ & $0.166129$ \\ 
$9$ & $11.18603$ & $14$ & $0.189390$ & $0.161426$ & $0.027964$ & $0.131469$ & $0.162837$ \\ 
$10$ & $10.86168$ & $13$ & $0.185992$ & $0.159246$ & $0.026746$ & $0.132625$ & $0.160488$ \\ 
$11$ & $10.55288$ & $12$ & $0.183169$ & $0.157632$ & $0.025537$ & $0.133654$ & $0.158741$ \\ 
$12$ & $10.27006$ & $11$ & $0.180795$ & $0.156396$ & $0.024399$ & $0.134568$ & $0.157397$ \\ 
$13$ & $10.01485$ & $10$ & $0.178768$ & $0.155424$ & $0.023344$ & $0.135381$ & $0.156336$ \\ 
$14$ & $9.78566$ & $9$ & $0.176996$ & $0.154642$ & $0.022354$ & $0.136108$ & $0.155480$ \\ 
$15$ & $9.57986$ & $9$ & $0.175421$ & $0.154002$ & $0.021419$ & $0.136759$ & $0.154776$ \\ 
$16$ & $9.39467$ & $8$ & $0.174048$ & $0.153470$ & $0.020579$ & $0.137347$ & $0.154190$ \\ 
$17$ & $9.22749$ & $8$ & $0.172805$ & $0.153021$ & $0.019784$ & $0.137879$ & $0.153694$ \\ 
$18$ & $9.07603$ & $7$ & $0.171664$ & $0.152639$ & $0.019025$ & $0.138363$ & $0.153270$ \\ 
$19$ & $8.93828$ & $7$ & $0.170679$ & $0.152309$ & $0.018370$ & $0.138804$ & $0.152904$ \\ 
$20$ & $8.81253$ & $7$ & $0.169756$ & $0.152023$ & $0.017733$ & $0.139209$ & $0.152586$ \\ 
$21$ & $8.69732$ & $7$ & $0.168892$ & $0.151772$ & $0.017120$ & $0.139580$ & $0.152306$ \\ 
$22$ & $8.59139$ & $6$ & $0.168101$ & $0.151551$ & $0.016549$ & $0.139924$ & $0.152059$ \\ 
$23$ & $8.49368$ & $6$ & $0.167411$ & $0.151355$ & $0.016056$ & $0.140241$ & $0.151839$ \\ 
$24$ & $8.40327$ & $6$ & $0.166758$ & $0.151180$ & $0.015578$ & $0.140535$ & $0.151642$ \\ 
$25$ & $8.31937$ & $6$ & $0.166141$ & $0.151023$ & $0.015118$ & $0.140810$ & $0.151465$ \\ 

\end{longtable}

In the $SU(2)$ case, the choice of the nucleus used to fix the energy scale of the dominant term $E_1$ through \rf{totalmassf}, i.e., the ratio $m_0/e_0$, does not significantly affect the resulting value of $m_0/e_0$. This is a consequence of isospin symmetry and of the fact that the nuclear binding energy is only of the order of $1\%$ of the total nuclear mass. These experimental facts ensure that the mass per baryon, $n_T \equiv E/A$, exhibits only a weak dependence on $A$, even for light nuclei. For convenience, one may therefore choose ${}^{1}$H to fix this ratio, that is, one may use $A=1$ and $m_A^{\rm exp.}=m_p$ on the r.h.s side of \rf{totalmassf}, where $m_p$ denotes the experimental proton mass, which leads to $m_0/e_0 = 1.95086\, MeV$.

\section{Aplication to nuclear physics for the case $G=SU(3)$}
\label{sec:su(3)app}
\setcounter{equation}{0}

The baryon number of the hypernuclei is given by $A = Z + N_n + n_Y$, where $Z$, $N_n$, and $n_Y$ are the numbers of protons, neutrons, and hyperons, respectively. However, from now on our discussion will be restricted to the physics of the $\Lambda$-hypernuclei, which possess only one hyperon, the $\Lambda$ particle, and so $n_Y = 1$. This choice is made because the separation energy, which is the energy necessary to separate the hyperon from the hypernucleus, is available for the $\Lambda$-hypernuclei over a wide range of mass numbers. For a $\Lambda$-hypernucleus the separation energy is given by
\be
B_\Lambda \equiv B\({}^{A}_\Lambda X\) - B\({}^{A-1}X\) \lab{separate}
\ee
where $B(\cdot)$ represents the binding energy of a $\Lambda$-hypernucleus ${}^{A}_{\Lambda}X$ or of the nucleus ${}^{A-1}X$, and $X$ denotes the chemical symbol of the element. On the other hand, the binding energy of a $\Lambda$-hypernucleus is, by definition,
\be
B\({}^{A}_\Lambda X\) = Z\,m_p+N_n\, m_n+m_{\Lambda}- m_T\({}^{A}_\Lambda X\) \lab{bindingh} 
\ee
where $m_T({}^{A}_{\Lambda}X)$ is the total mass of the hypernucleus ${}^{A}_{\Lambda}X$, $m_{\Lambda}$ is the $\Lambda$-hyperon mass, and $m_p$ and $m_n$ are the proton and neutron masses, respectively. Consequently, once the separation energy $B_{\Lambda}$ of the $\Lambda$-hypernucleus and the binding energy per nucleon $B({}^{A-1}X)$ of the core nucleus are known, one obtains the total binding energy $B({}^{A}_{\Lambda}X) = E_B^{{\rm exp.}}({}^{A-1}X)(A-1) + B_{\Lambda}$, where the experimental values of $E_B({}^{A-1}X)$ are known experimentally for a wide range of nuclei. Since experimentally $B_{\Lambda}>0$, the addition of a $\Lambda$ hyperon increases the total binding energy. Using also \rf{bindingh}, together with $N_n = A - Z - 1$ and $\Delta_p \equiv m_n - m_p$, the total experimental binding energy per baryon, $E_B^{{\rm exp.}}({}^{A}_{\Lambda}X) \equiv B({}^{A}_{\Lambda}X)/A$, and the experimental mass per baryon $n_T^{{\rm exp.}} \equiv m_T/A$ are given, respectively, by
\br
E_B^{{\rm exp.}}\({}^{A}_\Lambda X\) &=&  \(1-\frac{1}{A}\)\,E_B^{{\rm exp.}}\({}^{A-1} X\) + \frac{B_\Lambda}{A}\, \lab{hyperbinding}\\
n_T^{{\rm exp.}}\({}^{A}_\Lambda X\)&=&\(1-\frac{1}{A}\)\,m_p+\(1-\frac{1}{A}-\frac{Z}{A}\)\,\Delta_p +\frac{m_{\Lambda}}{A}- E_B^{{\rm exp.}}\({}^{A}_\Lambda X\) \lab{hypermass}
\er
where $m_p=938.27208816\, MeV$, $m_n=939.5654205\,MeV$ and $m_\Lambda = 1115.683\, MeV$, according to \cite{ParticleDataGroup:2024cfk}.

The binding energy of nuclei can be approximated by a semi-empirical mass formula called the Bethe--Weizsäcker (BW) mass formula, which possesses a modification, denoted by BWM, that improves the description of light (non-strange) nuclei, covering the spectra from $Z=3$ to $Z=83$. The binding energy of hypernuclei can be obtained from a semi-empirical extension of the BWM formula that considers a nucleus bound to one or more hyperons, the so-called BWMH formula, given by \cite{Samanta:2005kd}
\br
B_{{BWMH}} &=& B_{{BWM}} + B_S; \qquad\qquad B_S = n_Y\,\left[c_0\,m_Y-c_1 -c_2\,\mid S\mid A^{-\frac{2}{3}}\right] \lab{BWMH}\\
B_{{BWM}} & = & a_v\,A-a_s\,A^{\frac{2}{3}}-a_c \,Z_T\(Z_T-1\)\,A^{-\frac{1}{3}}-a_{sym}\,\frac{\(N_n-Z\)^2}{\(1+e^{-\frac{A}{k}}\)\,A} +\(1-e^{-\frac{A}{c_3}}\)\,\delta \nonumber
\er
where the term $B_{{\rm BWM}}$ corresponds with the BWM formula, $Z_T = Z+n_Y\,q_Y$ with $q$ being the charge number with proper sign of the hyperon, and $\delta= 0$, if $N_n+Z$ is odd, $\delta= 12\,MeV/\sqrt{A}$ if $Z$ and $N_n$ are even and $\delta= -12\,MeV/\sqrt{A}$ if $Z$ and $N_n$ are odd. For the $\Lambda$-hypernuclei one has $n_Y=1$ and $q=0$, since $\Lambda$ has zero electric charge. In addition, $c_3=30$, $k=17$, $a_v = 15.777\, MeV$, $a_s = 18.34\, MeV$, $a_c = 0.71\, MeV$, and  $a_{sym} = 23.21\, MeV$.

Considering the list of $\Lambda$-hypernuclei given in Table \ref{tableH}, the experimental data show that the values of $B_\Lambda/A$ become small compared with the binding energy per baryon for heavy hypernuclei, decreasing monotonically with $A$ from $0.39020\,MeV$ at $A=51$ to $0.12740\,MeV$ at $A=208$. Therefore, the separation energy provides only a small contribution to the binding energy per baryon of heavy hypernuclei with $A \geq 51$.

In this work we adopt the approximation that the baryon density of $\Lambda$-hypernuclei exhibits the same asymptotic decay rate, $a = 0.524\,\mathrm{fm}$, as that of ordinary nuclei, as used in Section \ref{sec:su(2)app}. In addition, we approximate the rms radius of the theoretical $A=56$ $\Lambda$-hypernucleus by the rms radius of ${}^{56}\mathrm{Fe}$. Consequently, the rms radii of all hypernuclei are approximated by the rms radii of nuclei with the same mass number $A$, because the choice of a single reference nucleus fixes all values of $\gamma$, as explained in Section \ref{sec:su(2)app}. In addition, throughout this section we set $s=3$ and $\kappa=25$.

The values of $\sigma_2$ and $\sigma_G$ are obtained from the fiting of the binding energy per baryon of the heavy hypernuclei with $A\geq 51$, which in our list of $31$ hypernuclei given in Table \ref{tableH}, corresponds with the four $\Lambda$-hypernuclei $^{51}_\Lambda$V, $^{89}_\Lambda$Y, $^{139}_\Lambda$La and  $^{208}_\Lambda$Pb. This leads to $\sigma_2 = 2.40334\,MeV$  and $\sigma_G= 9.78134 \times 10^{-3}$, which a RMSD of the binding energy per baryon of $\Delta\,E_B^{A \geq 51}= 0.03694\,MeV$.

\begin{figure}[htp!]
\begin{center}
		\includegraphics[scale=0.47]{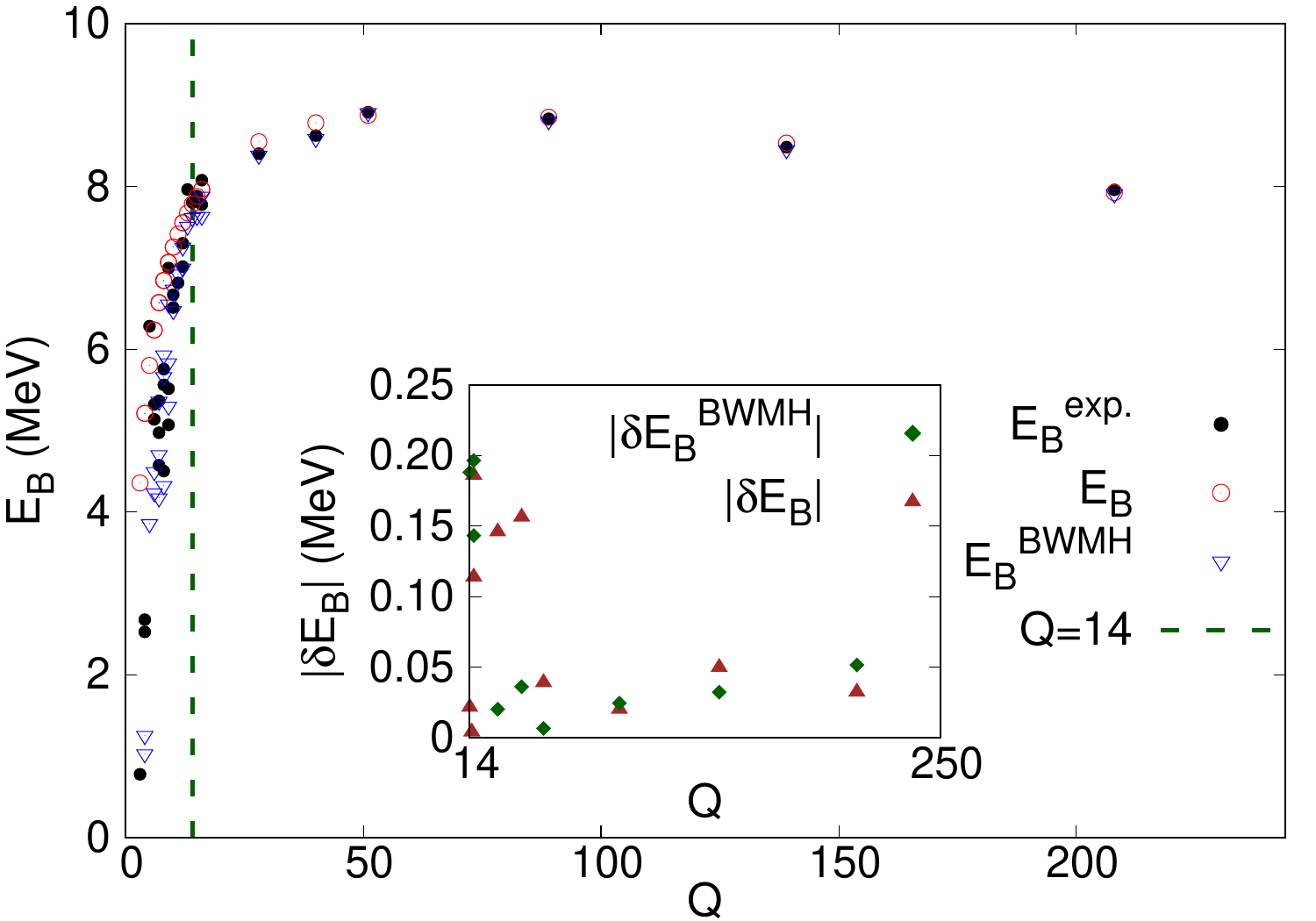}
		\caption{The binding energy per baryon $E_B$ and $\delta E_B$ as a function of $Q$ for $s=3$ and $\kappa=25$, corresponding to the minima of $\Delta E_B^{A\geq 51}$ given by $\sigma_2=2.40334\, MeV$ and $\sigma_G=9.78134 \times 10^{-3}$ with $\Delta\,E_B^{A \geq 51}= 0.03694\,MeV$.}
		\label{fig:bindingsu3}
\end{center} 
\end{figure}

Figure \ref{fig:bindingsu3} and Table \ref{tableH} show the values of the theoretical and experimental binding energy per baryon, obtained respectively through \rf{binding} and \rf{hyperbinding}. In the region $A\geq 14$, the maximum value of the error of the theoretical binding energy in relation to its experimental value, i.e. the ratio $\mid 1-E_B/E_B^{{\rm exp.}}\mid $, corresponds to $2.39\%$ for $^{16}_\Lambda$O, showing that as long as the $\Lambda$-hypernuclei is heavy enough, the deviation of the binding energy from experiment remains small. The values of $ \delta E_B \equiv \mid E_B- E_B^{{\rm exp.}}\mid$ and $ \delta E_B^{{\rm BWMH}} \equiv \mid E_B^{{\rm BWMH}}- E_B^{{\rm exp.}}\mid$ also are show in Figure \ref{fig:bindingsu3} for $A\geq 14$, where $E_B^{{\rm BWMH}} \equiv B_{{\rm BWMH}}/A$. The BWMH semi-empirical formula and the Generalized False Vacuum Skyrme model provide similar results as long as $A$ is large enough, with both $\delta E_B$ and $E_B^{{\rm BWMH}}$ being lower than $0.2\,MeV$ in the region $A\geq 14$. In fact, in the very heavy regime $A\geq 51$, the maximum value of the difference $\mid E_B- E_B^{{\rm BWMH}}\mid$ is $0.04470\,MeV$ at $^{89}_\Lambda$Y, which corresponds to only $0.51\%$ of its experimental binding energy per baryon. In this region, the results for the binding energy obtained from our model become more accurate, with the maximum value of $\mid 1 - E_B/E_B^{{\rm exp.}}\mid$ equal to $0.59\%$ for $^{139}_\Lambda\mathrm{La}$. This represents a reduction by more than a factor of four compared with the previous region.


Although for $(A,Z) = (1,0)$ one obtains from \rf{hypermass} that $n_T = m_\Lambda$, for sufficiently large $A$ the contribution of $m_\Lambda/A$ to $n_T$ becomes small in comparison with the proton mass. Indeed, the experimental values of the ratio $m_\Lambda/(m_p\, A)$ decrease from $0.0233$ at $A=51$ to $0.0057$ at $A=208$. Since the $\Lambda$-hyperon is approximately $18.91\%$ heavier than the proton, in light $\Lambda$-hypernuclei the addition or removal of a single nucleon produces a substantial change in the mass per baryon $n_T$. The same occurs when a $\Lambda$-hyperon is replaced by a proton or a neutron.

A fluid model in which the energy depends on $A$, such as the Generalized False Vacuum Skyrme model, may be applied, at least at the classical level, in the regime where $n_T$ varies slowly with $A$. Therefore, one may fix the ratio $m_0/e_0$ through \rf{totalmassf} by using the very heavy $\Lambda$-hypernucleus ${}^{208}_{\Lambda}\mathrm{Pb}$, for which $A=208$ and $m_{208}^{{\rm exp.}} = n_T^{{\rm exp.}}({}^{208}_{\Lambda}\mathrm{Pb})\,208$, with the value of $n_T^{{\rm exp.}}({}^{208}_{\Lambda}\mathrm{Pb})$ obtained from \rf{hypermass}, leading to $m_0/e_0=1.95354\,MeV$. This completes the determination of the coupling constants described in Section \ref{sec:fixing}. For $A\geq 51$, the error of $n_T$ in relation to its experimental value $n_T^{{\rm exp.}}$, denoted by $\Delta n= \mid 1.0-n_T/n_T^{{\rm exp.}}\mid $, is respectively $0.27\%,\, 0.12\%,\, 0.05\%, \,0\%$ at $A=51,\,89,\, 139,\,208$, see Table \ref{tableH}, while the mass gap between the theoretical and experimental mass per baryon $\mid n_T-n_T^{{\rm exp.}}\mid$ corresponds to $28.3\%,\,12.8\%,\,5.7\%,\,0\%$ of the experimental binding energy per baryon $E_B^{{\rm exp.}}$. 

\begin{longtable}{@{\extracolsep{\fill}}c c c c c c c c c c@{}}
\caption{The experimental values of the separation energy $B_\Lambda$ of a set of 31 $\Lambda$-hypernuclei, with the corresponding references, the experimental binding energy per baryon $E_B^{\rm exp.}$ obtained from \rf{hyperbinding}, where the values of $E_B^{\rm exp.}\({}^{A-1} X\)$ are given in \cite{ame2016}, the values of $E_B$ for $s=3$ and $\kappa=25$ obtained by using the coupling constant fixing described in this Section, and the values of $E_B^{{\rm BWMH}} \equiv B_{{\rm BWMH}}/A$ obtained using \rf{BWMH}, all in $MeV$ units. In addition, the total theoretical mass per baryon $n_T = E/A$, where $E$ is obtained from \rf{energyf}, its experimental values $n_T^{{\rm exp.}}$ obtained from \rf{hypermass}, and the modulus of their difference $\Delta n \equiv \mid 1-n_T/n_T^{{\rm exp.}}\mid $ in percent, all in proton-mass ($m_p$) units, are also presented.} \label{tableH} \\

\hline
$Z$ & $A$ & $B_\Lambda$ & $E_B^{{\rm exp.}} $ & $E_B$ & $E_B^{{\rm BWMH}}$ & $n_T^{{\rm exp.}}$ & $n_T$ & $\Delta n$ & Name \\
\hline
\endfirsthead

\hline
$Z$ & $A$ & $B_\Lambda$ & $E_B^{{\rm exp.}} $ & $E_B$ & $E_B^{{\rm BWMH}}$ & $n_T^{{\rm exp.}}$ & $n_T$ & $\Delta n$ & Name \\
\hline
\endhead

\hline
\endfoot

\hline
\endlastfoot
 
$1$ & $3$ & $0.102$ \cite{1} & $0.77552$ & $4.35698$ & $-1.40473$ & $1.06266$ & $0.99706$ & $6.17329$ & $^{3}_\Lambda$H \\ 
$1$ & $4$ & $2.22$ \cite{2} & $2.67545$ & $5.21312$ & $1.25047$ & $1.04511$ & $0.99615$ & $4.68482$ & $^{4}_\Lambda$H \\ 
$2$ & $4$ & $2.38$ \cite{2} & $2.52451$ & $5.21312$ & $1.02683$ & $1.04492$ & $0.99615$ & $4.66806$ & $^{4}_\Lambda$He \\ 
$2$ & $5$ & $3.12$ \cite{3} & $6.28313$ & $5.80151$ & $3.85443$ & $1.03167$ & $0.99552$ & $3.50418$ & $^{5}_\Lambda$He \\ 
$2$ & $6$ & $4.38$ \cite{4} & $5.32333$ & $6.23644$ & $4.49627$ & $1.02653$ & $0.99506$ & $3.06598$ & $^{6}_\Lambda$He \\ 
$3$ & $6$ & $4.5$ \cite{77} & $5.13833$ & $6.23644$ & $4.23579$ & $1.02650$ & $0.99506$ & $3.06290$ & $^{6}_\Lambda$Li \\ 
$2$ & $7$ & $5.55$ \cite{5} & $4.97444$ & $6.57390$ & $4.70257$ & $1.02250$ & $0.99470$ & $2.71895$ & $^{7}_\Lambda$He \\ 
$3$ & $7$ & $5.58$ \cite{8} & $5.36771$ & $6.57390$ & $5.36048$ & $1.02188$ & $0.99470$ & $2.66030$ & $^{7}_\Lambda$Li \\ 
$4$ & $7$ & $5.09$ \cite{3} & $4.57331$ & $6.57390$ & $4.17235$ & $1.02253$ & $0.99470$ & $2.72215$ & $^{7}_\Lambda$Be \\ 
$2$ & $8$ & $7.16$ \cite{6} & $4.50271$ & $6.84495$ & $4.32179$ & $1.01970$ & $0.99441$ & $2.48016$ & $^{8}_\Lambda$He \\ 
$3$ & $8$ & $6.81$ \cite{3} & $5.75688$ & $6.84495$ & $5.93007$ & $1.01819$ & $0.99441$ & $2.33564$ & $^{8}_\Lambda$Li \\ 
$4$ & $8$ & $6.91$ \cite{3} & $5.56385$ & $6.84495$ & $5.66382$ & $1.01822$ & $0.99441$ & $2.33884$ & $^{8}_\Lambda$Be \\ 
$3$ & $9$ & $8.36$ \cite{7} & $5.51530$ & $7.06833$ & $5.83150$ & $1.01590$ & $0.99417$ & $2.13872$ & $^{9}_\Lambda$Li \\ 
$4$ & $9$ & $6.49$ \cite{9} & $6.99883$ & $7.06833$ & $6.55566$ & $1.01416$ & $0.99417$ & $1.97137$ & $^{9}_\Lambda$Be \\ 
$5$ & $9$ & $7.89$ \cite{3} & $5.06969$ & $7.06833$ & $5.30054$ & $1.01607$ & $0.99417$ & $2.15496$ & $^{9}_\Lambda$B \\ 
$4$ & $10$ & $8.55$ \cite{10} & $6.67140$ & $7.25614$ & $6.73796$ & $1.01249$ & $0.99397$ & $1.82893$ & $^{10}_\Lambda$Be \\ 
$5$ & $10$ & $8.82$ \cite{3} & $6.51336$ & $7.25614$ & $6.47432$ & $1.01252$ & $0.99397$ & $1.83190$ & $^{10}_\Lambda$B \\ 
$5$ & $11$ & $10.24$ \cite{3} & $6.81735$ & $7.41657$ & $6.92449$ & $1.01055$ & $0.99380$ & $1.65766$ & $^{11}_\Lambda$B \\ 
$5$ & $12$ & $11.45$ \cite{3} & $7.30459$ & $7.55539$ & $7.25683$ & $1.00866$ & $0.99365$ & $1.48815$ & $^{12}_\Lambda$B \\ 
$6$ & $12$ & $10.75$ \cite{9} & $7.01592$ & $7.55539$ & $6.99840$ & $1.00885$ & $0.99365$ & $1.50698$ & $^{12}_\Lambda$C \\ 
$6$ & $13$ & $11.39$ \cite{13} & $7.96552$ & $7.67679$ & $7.51399$ & $1.00669$ & $0.99352$ & $1.30827$ & $^{13}_\Lambda$C \\ 
$6$ & $14$ & $12.17$ \cite{12} & $7.80557$ & $7.78391$ & $7.61761$ & $1.00588$ & $0.99341$ & $1.23962$ & $^{14}_\Lambda$C \\ 
$7$ & $15$ & $13.59$ \cite{12} & $7.88324$ & $7.87914$ & $7.63102$ & $1.00485$ & $0.99331$ & $1.14857$ & $^{15}_\Lambda$N \\ 
$7$ & $16$ & $13.76$ \cite{14} & $8.07824$ & $7.96437$ & $7.88177$ & $1.00390$ & $0.99321$ & $1.06410$ & $^{16}_\Lambda$N \\ 
$8$ & $16$ & $12.50$ \cite{9} & $7.77846$ & $7.96437$ & $7.63523$ & $1.00413$ & $0.99321$ & $1.08710$ & $^{16}_\Lambda$O \\ 
$14$ & $28$ & $16.00$ \cite{9} & $8.40561$ & $8.55145$ & $8.38535$ & $0.99843$ & $0.99259$ & $0.58545$ & $^{28}_\Lambda$Si \\ 
$20$ & $40$ & $18.70$ \cite{9} & $8.62793$ & $8.78430$ & $8.59164$ & $0.99619$ & $0.99234$ & $0.38602$ & $^{40}_\Lambda$Ca \\ 
$23$ & $51$ & $19.9$ \cite{9} & $8.91561$ & $8.87650$ & $8.90892$ & $0.99494$ & $0.99224$ & $0.27063$ & $^{51}_\Lambda$V \\ 
$39$ & $89$ & $22.1$ \cite{9} & $8.83330$ & $8.85342$ & $8.80873$ & $0.99347$ & $0.99227$ & $0.12098$ & $^{89}_\Lambda$Y \\ 
$57$ & $139$ & $23.8$ \cite{16} & $8.48610$ & $8.53589$ & $8.45376$ & $0.99312$ & $0.99261$ & $0.05172$ & $^{139}_\Lambda$La \\ 
$82$ & $208$ & $26.5$ \cite{16} & $7.95943$ & $7.92713$ & $7.90782$ & $0.99325$ & $0.99325$ & $0.00000$ & $^{208}_\Lambda$Pb \\ 
\end{longtable}

\begin{figure}[htp!]
\begin{center}
		\includegraphics[scale=0.47]{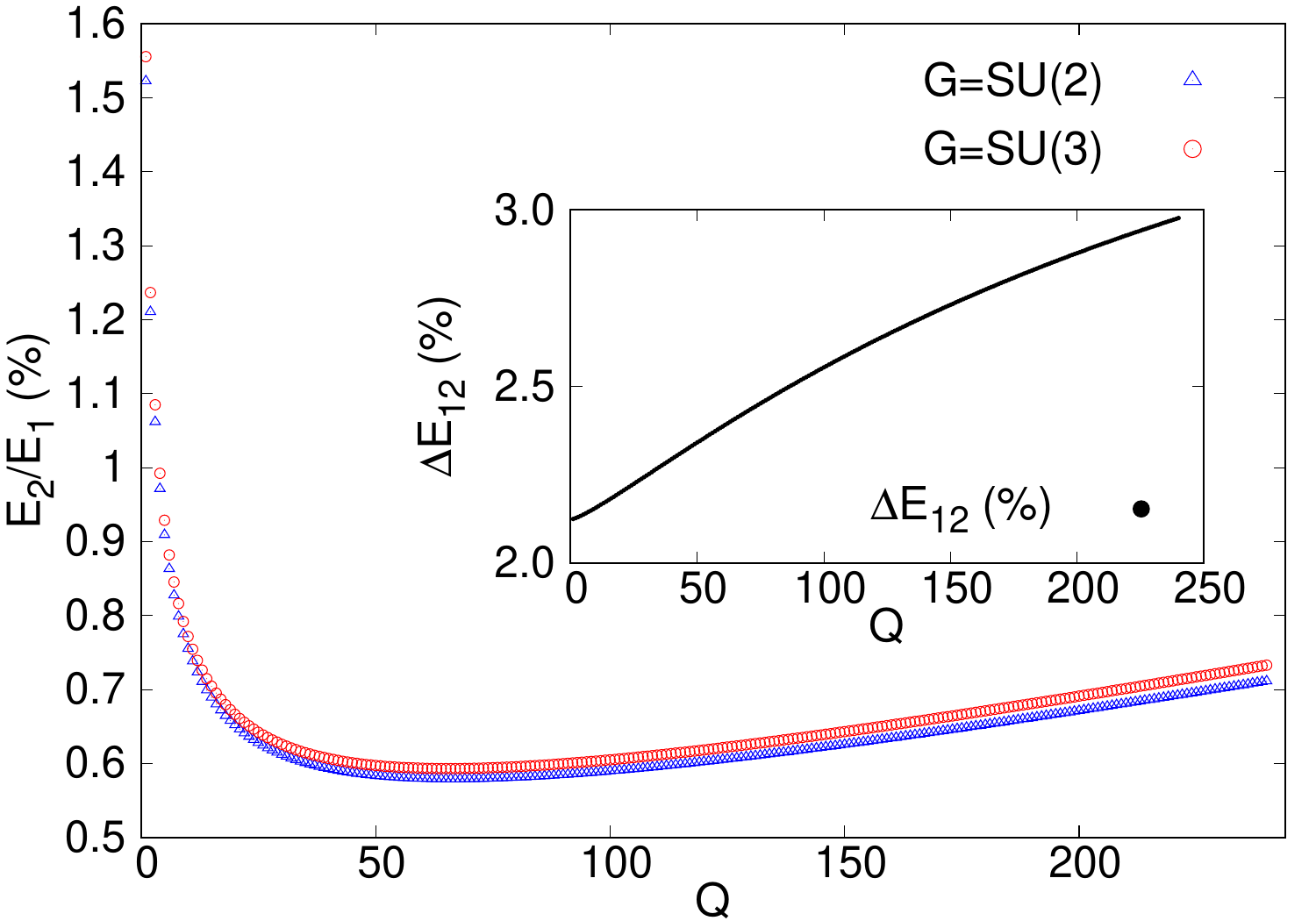}
\caption{The theoretical ratio $E_2/E_1$ in percent for $s=3$ and $\kappa=25$ for $G=SU(2)$ and $G=SU(3)$, with the coupling constants fixed as described in Sections \ref{sec:su(2)app} and \ref{sec:su(3)app}, respectively, together with the values of $\Delta E_{12} \equiv \(1-\frac{\(E_2/E_1\)^{G=SU(2)}}{\(E_2/E_1\)^{G=SU(3)}}\)$, also expressed in percent.}
		\label{fig:e2e1}
\end{center} 
\end{figure}

Figure \ref{fig:e2e1} shows the ratio $E_2/E_1$ for the $G=SU(2)$ and $G=SU(3)$ cases with $s=3$ and $\kappa=25$, where the coupling constants are fixed using the approach given in Sections \ref{sec:su(2)app} and \ref{sec:su(3)app}, respectively, and the quantity $\Delta E_{12} \equiv \(1-\frac{\(E_2/E_1\)^{G=SU(2)}}{\(E_2/E_1\)^{G=SU(3)}}\)$. Clearly, $E_2$ is two orders of magnitude lower than $E_1$ in both cases, showing that the model possesses a two–energy scale regime. In fact, the highest ratio $E_2/E_1$ occurs at $Q=1$ and corresponds to $1.5226\%$ for $G=SU(2)$ and $1.5557\%$ for $G=SU(3)$, and $E_2$ is lower than $1\%$ of $E_1$ for $Q \geq 4$ in both cases. In addition, $\(E_2/E_1\)^{G=SU(2)}$ is slightly higher, by approximately $2$--$3\%$, than ${\(E_2/E_1\)^{G=SU(3)}}$ for each value of $Q$.

\begin{figure}[htp!]
\begin{center}
		\includegraphics[scale=0.47]{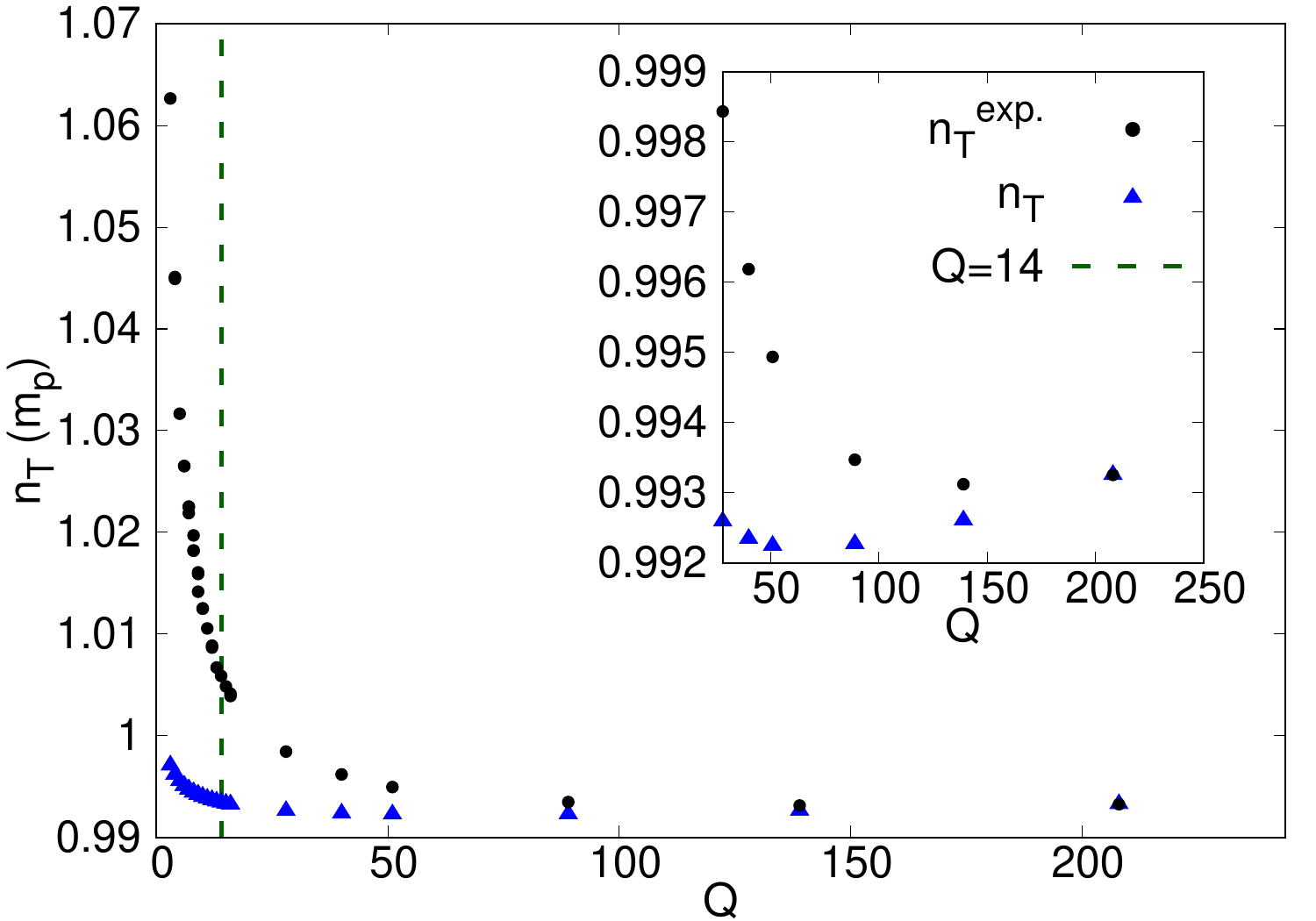}
      	\caption{The experimental mass per baryon $n_T^{{\rm exp.}}$ obtained through \rf{hypermass} and the theoretical mass per baryon $n_T$ for $s=3$ and $\kappa=25$, with the coupling constants fixed as described in Sections \ref{sec:su(3)app}, in proton mass ($m_p$) units.}
		\label{fig:masssu3}
\end{center} 
\end{figure}

\section{Conclusion}
\label{sec:conclusion}
\setcounter{equation}{0}

Our generalization of the False Vacuum Skyrme model, for any simple compact Lie group $G$ that leads to Hermitian symmetric spaces, largely extends the results of the original $G=SU(2)$ model. The global minima of the energy of the generalized model \rf{false0} can be attained by any $h$ fields satisfying the self-duality equations \rf{self} of the generalized BPS Skyrme model \rf{skyrmebps}, and the Skyrme field that minimizes $E_2$. This Skyrme field must lead to a spherically symmetric topological charge density ${\cal Q}$, since the minimization of $E_2$ falls into the well-known reduced problem defined in the context of the false vacuum decay.

We use the generalized rational map ansatz to study how to impose the spherical symmetry over ${\cal Q}$, which can be achieved by imposing the condition \rf{condition} over the Skyrme field, or more precisely over the $S$ matrix \rf{holog}, where $F(z,\,\bar{z})={\rm const.}>0$. A crucial consequence of such a condition is that the total energy density of such field configurations, up to a topological term, also has spherical symmetry. 

The total energy density of the global energy minimizers of the generalized False Vacuum Skyrme model corresponds, using the condition \rf{condition}, to the same functional energy of the case $G=SU(2)$, on top of the rescaling of the coupling constant $\beta_G$ that happens for $F \neq 1$. Once this condition is satisfied, the shape of the binding energy, which grows monotonically, saturates, and falls slowly for sufficiently large $Q$, the two energy scale regime of the total energy, and the rms radius, which grows as $Q^{\frac{1}{3}}$ near the critical value of the integration constants, are preserved. The model behaves as a classical field realization of the well-known phenomenological liquid drop model and may have other applications beyond nuclei.

In the case of the $G=SU(2)$ application to the description of the nuclei, our numerical results show that the predicted baryon density is in better agreement with the experimental data for large values of $\kappa$. Indeed, the error in the nuclear density at the centre of the nucleus for $^{208}$Pb, relative to its experimental value given in \cite{prex}, is less than or equal to $3.10\%$ as long as $\kappa \geq 20$. Furthermore, for $A \geq 12$, the model still provides an accurate description of both the RMS radius and the binding energy per nucleon. Specifically, for $\kappa \geq 6$, we find that $\Delta R^{A \geq 12}$ and $\Delta E_B^{A \geq 12}$ are approximately $0.04 \,fm$ and $0.04 \,MeV$, respectively. This significantly extends the knowledge about the optimal choice of the potential in the case of $G=SU(2)$ to reproduce the experimental data.

In the case of the $G=SU(3)$ application to the description of the $\Lambda$-hypernuclei, despite the loss of isospin symmetry, our model provides a reasonable description of the binding energies of the $\Lambda$-hypernuclei and their masses, as long as the baryon number is sufficiently large. Choosing $\kappa=25$ and $s=3$, the binding energy has an error equal to or less than $5.22\%$ for $A\geq 6$, and in the very heavy region $A\geq 51$ it drops to less than half. The error in the total mass decreases monotonically from $0.271\%$ at $A=51$ to zero at $A=208$ in our list of $31$ $\Lambda$-hypernuclei. However, for very light nuclei, the error in the total mass is very large in relation to the binding energy due to the significant difference between the masses of the proton or neutron and the $\Lambda$-hyperon. 

For $G=SU(p+q)$, with $p$ and $q$ being positive integers, which leads to the Hermitian symmetric space $SU(p+q)/SU(p) \otimes SU(q) \otimes U(1)$, we explicitly construct Skyrme field that satisfy the condition \rf{condition} with $F=1$. The total energy \rf{false0} and the topological charge density given in \rf{psidef} reduce completely to the $G=SU(2)$ case, in the sense that even all the coupling constants and the profile function $f(r)$ are the exactly the same for each value of $Q$. Although for different values of $N = p + q$, which in the usual Skyrme model is interpreted as the number of light quark flavours, we may have to use other parameters to fix the coupling constants, as we have done for the $G=SU(3)$ case, the form of the binding energy, rms radius, and topological charge density remains the same. This major result shows that this topological model may possess other physical applications for groups larger than $SU(2)$ in regimes where the most relevant degree of freedom is the baryonic density.

This work deepens our understanding of how effective classical field theories may play a role in the description of nuclear matter.  It also opens a path to include, in the $G=SU(2)$ case, new extra terms to enrich the energy spectrum of the model, especially for very light nucleons with mass number $A < 12$. In this regime, the results for the spectrum are not as good when compared with the experimental data, which is expected, since the model imposes spherical symmetry on the energy and topological charge densities. 

One possible direction for improving the spectrum of our model in the case $G = SU(N)$ is to study alternative matrices $S$ inside in the holomorphic map ansatz, given by \rf{Spq}, for $N \geq 3$. This approach can lead to a spherically symmetric topological charge density for angular functions $F(z, \bar{z})$ beyond the standard case $F = 1$. However, for $N = 2$, spherical symmetry arises within the generalized rational map ansatz only when $F = 1$.

The breaking of the self-duality equations \rf{self} for the $h$ fields may destroy the spherically symmetric nature of the topological charge density. This can be achieved by introducing kinetic or potential terms for the $h$ fields. However, we must propose modifications that improve the spectrum of light nuclei (hypernuclei), while leading to very small corrections to the spectrum of heavy nuclei (hypernuclei), where the model gives an accurate description of the binding energy, which can be challenging. 
 
\vspace{2cm}

{\bf Acknowledgements:} LAF is supported by Conselho Nacional de Desenvolvimento Cient\'ifico e Tecnol\'ogico - CNPq (contract 307833/2022-4), and Funda\c c\~ao de Amparo \`a Pesquisa do Estado de S\~ao Paulo - FAPESP (contract 2022/00808-7). LRL is supported by the grant 2022/15107-4, São Paulo Research Foundation (FAPESP).


\providecommand{\href}[2]{#2}\begingroup\raggedright\endgroup

\end{document}